\documentclass[elsarticle-num]{elsarticle}
\pdfoutput=1

\usepackage{tikz}

\usepackage{amsmath}
\usepackage{amssymb}

\usepackage{natbib}

\usepackage{url}
\usepackage{deluxetable}

\newcommand{\sun}{\odot}

\begin{document}

\begin{frontmatter}
	\title{Propagation in 3D
spiral-arm cosmic-ray source distribution models and secondary particle production using \textsc{Picard}}

	\author[ibk_ap]{	R.~Kissmann\corref{cor_auth}}
	\ead{ralf.kissmann@uibk.ac.at}
	
	\author[ibk_ap]{	M.~Werner}
	\author[ibk_ap]{O.~Reimer}
	\author[MPIE]{A.~W.~Strong}
	
	\address[ibk_ap]{Institut f\"ur Astro- und Teilchenphysik,	  Leopold-Franzens-Universit\"at Innsbruck, A-6020 Innsbruck, Austria}
	  
	\address[MPIE]{Max Planck Institut f\"ur extraterrestrische Physik, Postfach 1312, D-85741 Garching, Germany}

	\cortext[cor_auth]{Corresponding author}

	\begin{abstract} We study the impact of  possible  spiral-arm distributions of Galactic cosmic-ray sources on the flux of various cosmic-ray nuclei throughout our Galaxy. We investigate model cosmic-ray spectra at the nominal position of the sun and 
at different positions within the Galaxy.	
The modelling is performed using the recently introduced numerical cosmic ray propagation code \textsc{Picard}. Assuming non-axisymmetric cosmic ray source distributions yields new insights on the behaviour of primary versus secondary nuclei.

 We find that primary cosmic rays are more strongly confined to the vicinity of the sources, while the distribution of secondary cosmic rays is much more homogeneous compared to the primaries.
This leads to stronger spatial variation in secondary to primary ratios when compared to axisymmetric source distribution models.
A good fit to the cosmic-ray data at Earth can be accomplished in different spiral-arm models, although leading to decisively different spatial distributions of the cosmic-ray flux. 
This results in very different cosmic ray anisotropies, where even a good fit to the data becomes possible.
Consequently, we advocate directions to seek best fit propagation parameters that take into account the higher complexity introduced by the spiral-arm structure on the cosmic-ray distribution.
We specifically investigate whether the flux at Earth is representative for a large fraction of the Galaxy. The variance among possible spiral-arm models allows us to quantify the spatial variation of the cosmic-ray flux within the Galaxy in presence of  non-axisymmetric   source distributions.
	\end{abstract}


\end{frontmatter}
	
\section{Introduction}
The majority of  Galactic cosmic-ray transport models assume our Galaxy to be azimuthally symmetric, thus allowing the solution of the transport problem efficiently in 2D cylindrical coordinates. The Milky Way, however, is known to be a spiral Galaxy (see, e.g. \cite{Vallee2014ApJS215_1}). Apart from the imprint on the gas distribution and the magnetic field (see, e.g. \cite{Jansson2012ApJ757_14, FerriereTerral2014AnA561_100}), the Galactic spiral arms are also thought to have an important impact on the distribution of the sources of Galactic cosmic rays (see, e.g. \cite{Shaviv2003NewA8_39, EffenbergerEtAl2012AnA547A120, BenyaminEtAl2014ApJ782_34}). This idea is motivated by cosmic-ray source candidates like pulsars and supernova remnants being rather young (see, e.g. \cite{Reynolds2008ARAnA46_89, BerezhkoVoelk1997APh7_183, EllisonEtAl2004AnA413_189, AmatoBlasi2006MNRAS371_1251}) and therefore mostly confined to the vicinity of the star-formation regions and thus to the vicinity of the spiral arms. 
Just recently, supernova remnants (SNRs) have received observational support as sources of cosmic rays (see \cite{AckermannEtAl2013Sci339_807}), supporting the link between cosmic-ray sources candidates and star-forming regions.

Nevertheless, in most of the Galactic cosmic-ray propagation models the sources are explicitly assumed to be axisymmetrically distributed in the Galaxy (see, e.g. \cite{StrongMoskalenko1998APJ509_212, AckermannEtAl2012ApJ750_3A}). Currently, however, we witness a transition in Galactic cosmic-ray modelling from two-dimensional azimuthally symmetric models to those that allow higher degrees of realism.
Recent studies of Galactic cosmic-ray propagation assuming a spiral-arm source distribution for the cosmic rays, e.g., include those by \citet{BenyaminEtAl2014ApJ782_34, EffenbergerEtAl2012AnA547A120, GaggeroEtAl2013PhRvL111_021102}; and \citet{WernerEtAl2015APh64_18}.

\citet{ShavivEtAl2009PhRvL103_1302} emphasized that an inhomogeneous source distribution should help to explain the observed increase in the  positron fraction (see \cite{Pamela2013PhRvL111_1102, AguilarEtAl2013PhRvL110_1102}). Correspondingly, \citet{GaggeroEtAl2013PhRvL111_021102} found that a spiral-arm source distribution can be used to explain the observed leptonic cosmic-ray spectra at Earth.
For this reason, additional primary sources of positrons were introduced.
Furthermore, \citet{BenyaminEtAl2014ApJ782_34} showed that the a spiral-arm source distribution can be used to explain the observed cosmic-ray Boron/Carbon  ratio (B/C) in a plain-diffusion model by taking the relative motion between Earth and the spiral arms into account.

These studies used a particular realisation for the spiral-arm source distribution each, where significant differences occur between the different models (see, e.g., the different spiral-arm models used in \cite{EffenbergerEtAl2012AnA547A120,BenyaminEtAl2014ApJ782_34}). These different choices reflect the different appearance of the Galactic spiral-arm structure in different tracers together with the problematic position of the observer within the Galaxy (see \cite{Steiman-CameronEtAl2010ApJ722_1460, Vallee2014ApJS215_1}).

This shows that, when exploring effects of a spiral-arm source distribution model, there is no unique such model. Properties of the spiral-arm model that  differ between the various studies include the number of the spiral arms, their pitch-angle, their width, and their tangent longitudes. Consequently, \citet{WernerEtAl2015APh64_18} systematically investigated the impact of different such source distributions on the model cosmic-ray spectra and the corresponding spatial distribution of hadronic and leptonic cosmic rays.

Here, we expand the study by \citet{WernerEtAl2015APh64_18} by taking the effects of the nuclear reaction network into account. Apart from demonstrating a fit of the secondary-to-primary ratios at the location of the sun, as is done in the models discussed abovem, also an investigation of the spatial variation of these quantities is warranted (see, e.g. \cite{Shaviv2003NewA8_39} for a discussion of the variation of $^{10}$Be/$^9$Be). This relates to the question, whether the cosmic-ray fluxes observed at the location of the sun are representative for other regions in the  Galaxy.

In the presence of spiral-arm cosmic-ray source distributions  the role of the propagation parameters needs to be re-addressed: can the standard parameters (see, e.g. \cite{TrottaEtAl2011ApJ729_106}) be adapted to allow a similarly good fit to the data in the presence of a spiral-arm source distribution, and 
what consequences arise for discrepancies?
 This also relates to the question of whether there are constraints from the cosmic-ray data that can be used to
 isolate a single model from the multitude of possible spiral-arm model realisations.
A problematic constraint is the observed cosmic ray anisotropy
in the 1\,TeV range, where propagation models using an axisymmetric
source distribution often predict a dipole amplitude of the anisotropy
about an order of magnitude higher than is observed (see \cite{StrongEtAl2007ARNPS57_285,EvoliEtAl2012PhRvL108_1102}).

We will address these points as follows. First we will briefly introduce the numerical framework used for the solution of the Galactic cosmic-ray propagation problem, where the validity of the scheme is demonstrated in the appendix.
Then we discuss our specific simulation setup in conjunction with the considered spiral-arm models.
  Subsequently, we present the results from different spiral-arm source distribution models.
Finally, we will conclude with a discussion of the consequences, where we demonstrate that the transition from axisymmetric to spatially three-dimensional modelling, including azimuthal variation, will allow to put further constraints on the cosmic-ray transport models, especially when combined with an analysis of the diffuse Galactic gamma-ray emission.

\section{Numerical Solution}
\label{SecNumSolution}
In this study we numerically compute the cosmic-ray flux in the Galaxy using the recently introduced code \textsc{Picard}. This code solves the cosmic-ray transport equation on a numerical grid with three spatial and one momentum dimension  (for details, see \cite{Kissmann2014APh55_37}). There, it was shown that \textsc{Picard} is particularly efficient in computing high resolution steady state solutions of the transport equation.

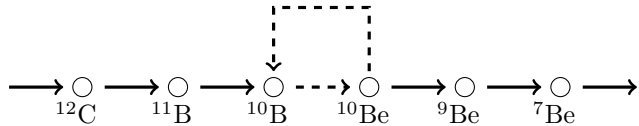
\begin{figure}
	\begin{center}
	\setlength{\unitlength}{0.0035\textwidth}
	\begin{tikzpicture}[x=\unitlength,y=\unitlength]
	\draw[very thick, ->, color=black] (-18,10) -- (-1,10);
	\draw[color=black] (5,10) circle [radius=3];	
	\node at (3,2) {$^{12}$C};
	\draw[very thick, ->, color=black] (12,10) -- (29,10);
	\draw[color=black] (35,10) circle [radius=3];
	\node at (33,2) {$^{11}$B};
	\draw[very thick, ->, color=black] (42,10) -- (59,10);
	\draw[color=black] (65,10) circle [radius=3];
	\node at (63,2) {$^{10}$B};
	\draw[very thick, ->, color=black, dashed] (72,10) -- (89,10);
	\draw[very thick, ->, color=black, dashed] (95,15) -- (95,35) -- (65,35) -- (65,15);
	\draw[color=black] (95,10) circle [radius=3];
	\node at (93,2) {$^{10}$Be};
	\draw[very thick, ->, color=black] (102,10) -- (119,10);
	\draw[color=black] (125,10) circle [radius=3];
	\node at (123,2) {$^{9}$Be};
	\draw[very thick, ->, color=black] (132,10) -- (149,10);
	\draw[color=black] (155,10) circle [radius=3];
	\node at (153,2) {$^{7}$Be};
	\draw[very thick, ->, color=black] (162,10) -- (179,10);
	\end{tikzpicture}
	\end{center}
	\caption{\label{FigIllustrationNucNet}Illustration of our handling of the nuclear network. In \textsc{Picard} the transport equation is only solved repeatedly in the range of the nuclear network, where a later (to the right in the image) nucleus can decay into an earlier one (to the left in the image). For more details see the text.}
\end{figure}

With regard to the implementation of the nuclear reaction network, which is essential to the present study, there are some important differences to other widely-used propagation codes like, e.g., \textsc{Galprop} (for the discussion of the nuclear network in the context of \textsc{Galprop} see, e.g. \cite{StrongMoskalenko2001AdSpR27_717, StrongMoskalenko2001ICRC5_1942}). Like in \textsc{Galprop} the transport equation is first solved for the heaviest nucleus $^{N}_{Z}$X in the nuclear reaction network. From the resulting distribution function the contribution of nucleus $^{N}_{Z}$X to the secondary spallation source term of all lighter particles can be computed. Subsequently the transport equation is solved for the isotope $^{N-1}_{Z}$X, where the solution proceeds to the element having charge $Z-1$  when all isotopes of the element with charge $Z$ have been addressed.

In this way all nuclei are ordered so  that most of them only decay (via spallation or radioactive decay) into nuclei which occur later in the network (for an illustration see Fig. \ref{FigIllustrationNucNet}). There are, however, a few nuclei that potentially produce secondaries that have already been treated earlier in the nuclear reaction network. Therefore, repeated runs of the entire network with a typical number of two network iterations (see e.g. \cite{StrongMoskalenko2001AdSpR27_717}) are required in \textsc{Galprop}.

We use a different strategy in \textsc{Picard}, where only parts of the network subject to the above effect are repeated. 
For this we first identify those nuclei that can decay into nuclei that appear earlier in the ordering scheme described above. Then only those parts of the network affected by this decay need to be solved repeatedly (for the specific example of the decay $^{10}$Be $\to$ $^{10}$B see Fig. \ref{FigIllustrationNucNet}, where the local network iteration is illustrated by the dashed arrows). This is much more efficient than a repetition of the entire network, thus allowing several repetitions of the potentially affected parts of the network with the same computational effort. \textsc{Galprop} uses the same time-integration scheme to obtain a solution for each network iteration, thus, leading to the same numerical cost for the computation of each network iteration. In contrast, the solver used in the \textsc{Picard} code detects whether a solution has been found, thus, allowing to compute a solution to the transport equation more quickly in repeated solutions for the same particle. In total the different implementation of the solution strategy of the nuclear reaction network results in a much higher efficiency in case of \textsc{Picard}, which also allows using a significantly higher number of network iterations. As discussed in \ref{AppNetIter}, in most cases two network iterations are sufficient to lead to a accurate solution for all relevant species. We will describe the specific setup that is used in the following.

\section{Propagation Models Setups}
\label{SecSetup}
Apart from the capability to efficiently handle spatially 3D transport problems, the new solver introduced in the \textsc{Picard} code offers the convenient feature to use the same transport parameters as the widely used \textsc{Galprop} code
(see \url{http://sourceforge.net/projects/galprop/}); we  use the same implementation of the nuclear cross-sections (see, e.g. \cite{StrongMoskalenko2001AdSpR27_717}) and of the various energy-loss processes (see, e.g. \cite{StrongMoskalenko1998APJ509_212}).

This allows us to use well-established propagation parameters sets. Furthermore, this enables us to adopt parameters addressed in 2D model setups to initialize 3D model realisations. In \ref{SecNumAnalysis} we document the validity of this approach quantitatively.

In this study we use the 2D reference models $^{\text{S}}$Y$^{\text{Z}}$4$^{\text{R}}$20$^{\text{T}}$150$^{\text{C}}$5, $^{\text{S}}$Y$^{\text{Z}}$4$^{\text{R}}$30$^{\text{T}}$150$^{\text{C}}$5, 
 $^{\text{S}}$Y$^{\text{Z}}$8$^{\text{R}}$20$^{\text{T}}$150$^{\text{C}}$5, and $^{\text{S}}$Y$^{\text{Z}}$8$^{\text{R}}$30$^{\text{T}}$150$^{\text{C}}$5 from \citet{AckermannEtAl2012ApJ750_3}, which will be referred to as models z4R20, z4R30, z8R20 and z8R30\footnote{Model z8R20 is only used in the investigation of the numerical setup in \ref{SecNumAnalysis}, where it was found to be not suitable for a 3D simulation setup.}, respectively.  This set of propagation parameters in these models was found to fit the cosmic-ray data, with the relevant transport parameters given in table \ref{TabParameters}.
 
With \textsc{Picard}, we solve models on a 3D Cartesian mesh in configuration space and a logarithmically equidistant mesh in momentum space.
Following the study of the discretisation error in \ref{AppendResStudy}, we use a resolution of 257 grid points in $x$ and $y$. For the $z$-dimension we used 65 grid points for the $z=4$\,kpc models and 129 grid points for the $z=8$\,kpc models. We cover the energy range from 0.01 to $10^6$\,GeV with 127 logarithmically equidistant grid points.

\begin{deluxetable}{lcccc}
\tabletypesize{\footnotesize}
\tablecolumns{5}
\tablewidth{0pt}
\tablecaption{\label{TabParameters}Transport parameters for all models used in this study. The source distribution used for a simulation is stated in the discussion in the main text. }
\tablehead{
\colhead{Parameters} & \colhead{Model 1} & \colhead{Model 2} & \colhead{Model 3} & \colhead{Model 4}\\
& \colhead{z4R20} & \colhead{z4R30} & \colhead{z8R20} & \colhead{z8R30}}
\startdata
\, Halo height [kpc] & 4 & 4 & 8 & 8\\
\, Galactic radius [kpc] & 20 & 30 & 20 & 30\\
\, Diffusion coefficient\tablenotemark{a} $D_0$ & 5.28 & 5.29 & 8.88 & 9.02 \\
\, Diffusion coefficient\tablenotemark{a} $\delta$ & 0.33 & 0.33 & 0.33 & 0.33 \\
\, $v_A$ [km s$^{-1}$] & 33.09 & 33.07 & 32.06 & 31.43 \\
Nuclei injection spectrum:\\
\, Index below break &	1.89 & 1.89 & 1.91 &  1.92\\
\, Index above break & 2.39 & 2.39 & 2.39 & 2.39 \\
\, Break energy [GeV] & 11.41 & 11.40 & 11.65 & 11.61\\
Normalisation:\\
\, Normalisation energy [GeV] & 108 & 108 & 108 & 108 \\
\, Normalisation flux\tablenotemark{b} & $4\cdot10^{-2}$ & $4\cdot10^{-2}$ & $4\cdot10^{-2}$ & $4\cdot10^{-2}$\\
\enddata
\vspace{-0.2cm}
\tablenotetext{a}{Diffusion coefficient is given as $D_{xx} = 10^{24} \beta D_0 (\rho / \rho_0)^{\delta}$ m$^2$\,s$^{-1}$ with $\rho_0 = 4$\,GV as reference rigidity.}
		\tablenotetext{b}{Normalisation flux given in units of [m$^{-2}$\,s$^{-1}$\,sr$^{-1}$\,(GeV/nucleon)$^{-1}$].}

\end{deluxetable}

For all axisymmetric models discussed in this study we used the source distribution based on the radial distribution of pulsars in the Galaxy by \citet{YusifovKuecuek2004AnA422_545} given as:
\begin{equation}
	\rho(r,z) =
	\left(\frac{r + r_{off}}{R_{\sun} + r_{off}}\right)^{\alpha}
	\exp \left\{\beta \left(\frac{r - R_{\sun}}{R_{\sun} + r_{off}}\right)\right\}
	\exp \left\{
	-\frac{|z|}{z_0}
	\right\}	
\end{equation}
with $R_{\sun} = 8.5$\,kpc, $r_{off} = 0.55$\,kpc, $z_0 = 0.2$\,kpc, $\alpha=1.64$, and $\beta = 4.01$. Additionally, we set $\rho(r,z) = 0$ when $r>15$\,kpc to have no cosmic-ray sources near the radial domain boundaries.

In the present study we investigate the influence of different spiral-arm cosmic-ray source distributions on the cosmic-ray flux within the Galaxy. The choice of such a source distribution is motivated by SNRs being one of the most relevant sources of cosmic rays. Since only relatively young SNRs seem to be able to accelerate cosmic rays to appreciable energies (see remark in the introduction in that regard),
most cosmic rays should be accelerated in those regions, where the supernovae originally occurred. These regions are the star-formation regions in the Galaxy, which are mostly found in the vicinity of the spiral arms and the Galactic bar. 

There exists a variety of different spiral-arm models for our Galaxy, and most assume four major spiral arms (see, e.g. \cite{Steiman-CameronEtAl2010ApJ722_1460, Vallee2014ApJS215_1}). However a sizeable fraction of models assume only two spiral arms  (but see \cite{Vallee2014MNRAS442_2993} for a discussion). This ambiguity is related to the position of the observer being located within the Galaxy. Apart from this, different studies use different tracers to identify the spiral arms, where models using evolved stars as tracers seem to prefer only two spiral arms (see \cite{Steiman-CameronEtAl2010ApJ722_1460, ChurchwellEtAl2009PASP121_213}),
possibly implying that the Galaxy only shows four spiral arms in the gas distribution but not in the distribution of old stellar populations (see \cite{Steiman-CameronEtAl2010ApJ722_1460}).
Furthermore, \citet{Vallee2014ApJS215_1} show that different tracers relate to different positions within the spiral arms, with cold molecular gas being found at the centre of the arm and hot dust at the outer rim.

The progenitors of SNRs are born in star-formation regions mostly related to spiral arms. Since the typical supernova explosion only occurs at the end of the lifetime of the progenitor star there might actually be a shift in time between the passage of a spiral arm and the average supernova, estimated to be about 15.4\,Myr (see \cite{Shaviv2003NewA8_39}). This time scale also translates into a shift between the majority of the active cosmic-ray sources and the observed position of the spiral arms. Additionally, \citet{Shaviv2003NewA8_39} use a broader distribution for the cosmic-ray sources than is implied by the spiral arms. This is motivated by the fact that not all supernovae happen at the same time, but are rather distributed in time according to the type of the progenitor star.

Consequently it is non-trivial to derive the appropriate source distribution from some spiral-arm model or related observations. On the one hand this allows for modification in the source model; on the other hand the related freedom means that a range of such source models should be applied to test the impact  on the resulting cosmic-ray flux at Earth. Therefore, we decided to investigate a subset of these models representative of as broad a range of spiral-arm parameters as possible.

As a representative for a four-arm model with logarithmic spiral arms we use a model taken from \citet{Steiman-CameronEtAl2010ApJ722_1460}, which was derived using observations of FIR cooling lines, [CII] and [NII], of the interstellar medium. These lines trace increased density and also UV radiation fields and are therefore thought to indicate the presence of star formation regions. For our implementation we use the parametrisation in Eq. (9) in \citet{Steiman-CameronEtAl2010ApJ722_1460} as described in \citet{WernerEtAl2015APh64_18}. The corresponding model is  referred to as the Steiman-model in this paper.

From the Steiman-model we also construct a two-arm model by only retaining the Scutum-Crux and the Perseus arms and adding a bar component with a half-length of 3.2\,kpc, with the corresponding parametrisation given in Eq. (4) in \citet{WernerEtAl2015APh64_18}. This model is qualitatively consistent with the barred two-arm model by \citet{ChurchwellEtAl2009PASP121_213}, which was based on the distribution of red-clump giant stars in the Galactic Legacy Infrared Mid-Plane Survey Extraordinaire (GLIMPSE)
 catalogue. Additionally, the model contains an extension of the Scutum-Crux arm in the outer Galaxy as found by \citet{DameThaddeus2011ApJ734L_24} in 21\,cm and CO observations. The two-arm model is  referred to as the Dame-model.

As a third model we use the NE2001-model which describes the distribution of free electrons in the Galaxy by \citet{CordesLazio2002astro_ph_7156}. This model was based on observation of pulsar dispersion measures. In this model the logarithmic spiral arms are locally modified and feature a different radial intensity dependence. Additionally, the model incorporates a nearby spiral-arm segment not present in the other two models.
 We used the authors spiral-arm configuration provided through \url{http://www.astro.cornell.edu/~cordes/NE2001/}.

These setups were chosen to have a wide variety of spiral-arm models to get an idea of their principal impact. The NE2001-model features broader spiral arms and thus a smoother source distribution than the other two models. At the same time the localisation of the sources is strongest in the Dame-model, making this two-arm model a rather extreme case, which seems actually to be contradicted by recent observations (see, e.g. \cite{UrquhartEtAl2014MNRAS437_1791}).

\section{Results from Cosmic-Ray Propagation Models}
In this sections, the impact of different source distributions is investigated and differences quantified. Differences are particular noteworthy in models with localized cosmic-ray sources, because primary cosmic rays are exclusively produced in the sources, whereas secondary cosmic rays are produced everywhere in the Galaxy by spallation reactions of heavier cosmic-ray nuclei with the interstellar gas. Still some concentration of the secondaries near the primary cosmic-ray sources can be expected since the density of primary cosmic rays is highest there. 

\begin{figure*}
	\setlength{\unitlength}{0.00044\textwidth}
	\begin{picture}(1100,938)(-100,-100)
	\put(360,-70){$x$ [kpc]}
	\put(-70,360){\rotatebox{90}{$y$ [kpc]}}
	\includegraphics[width=1000\unitlength]{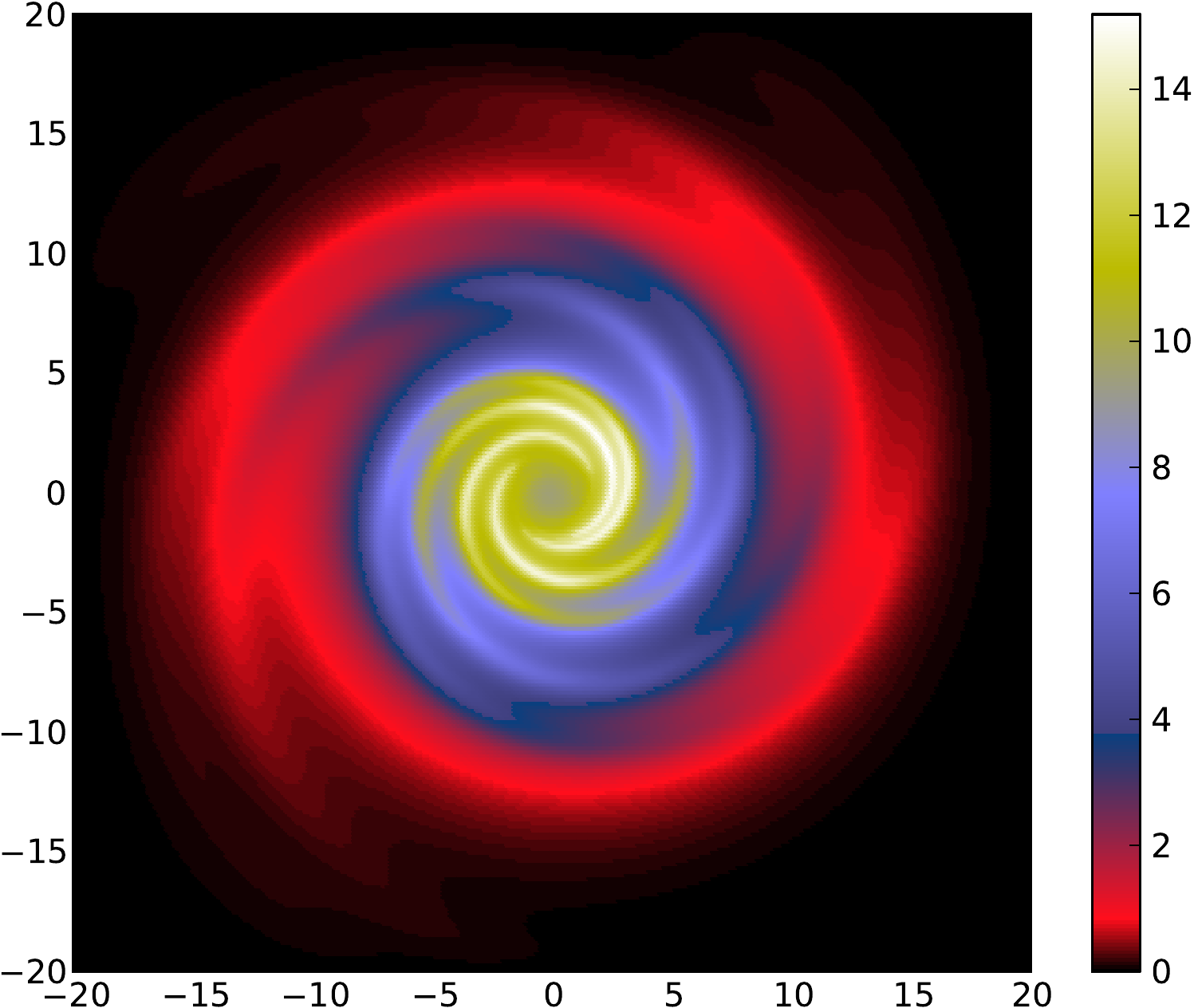}
	\end{picture}
	\hfill
	\begin{picture}(1000,938)(0,-100)
	\put(360,-70){$x$ [kpc]}
	\includegraphics[width=1000\unitlength]{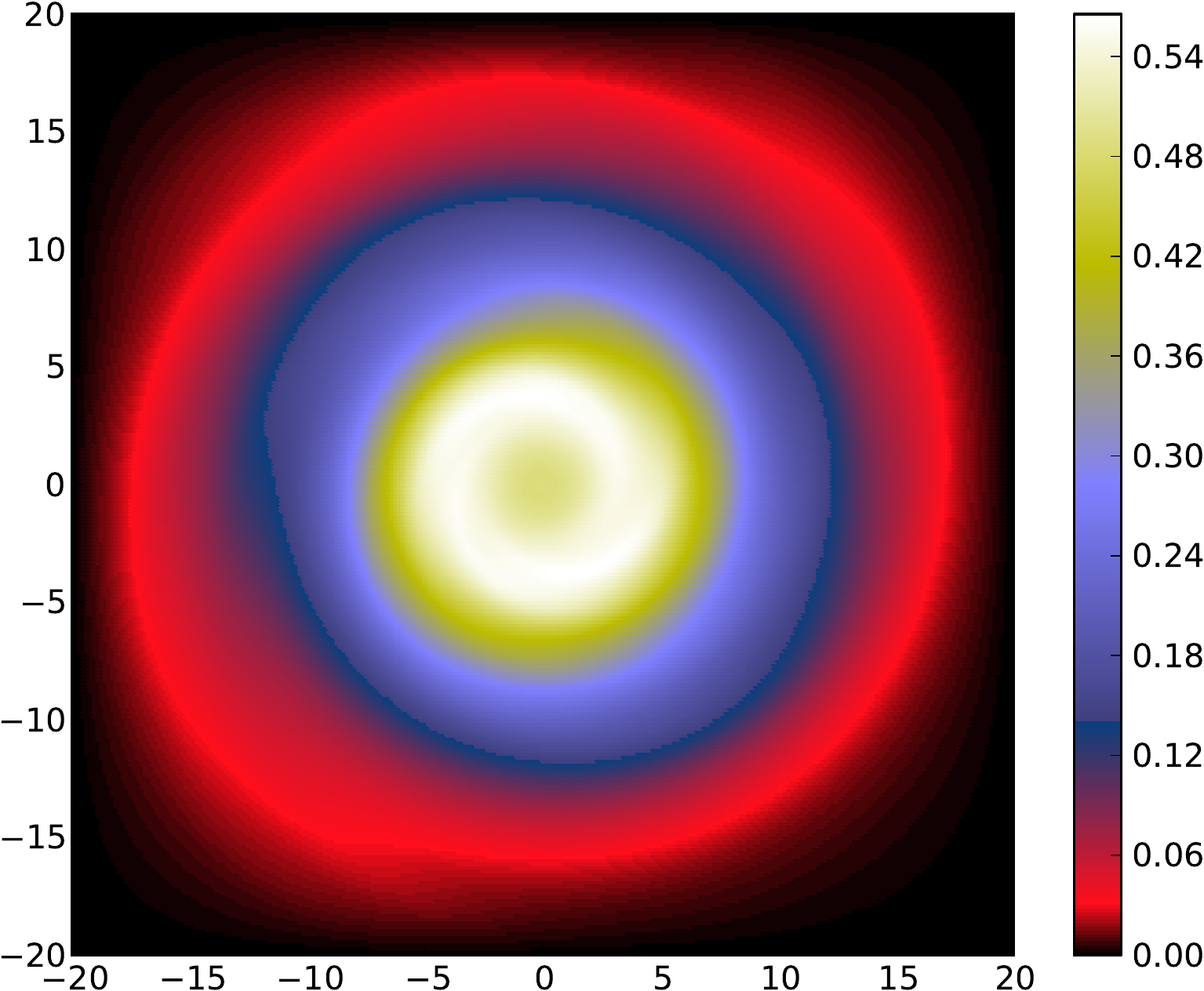}
	\end{picture}
	\caption{\label{FigDist_CarbonBoron_GalPlain}Cosmic-ray flux in the Galactic plane at an energy of $\sim$10\,GeV/nucleon for the Steiman source distribution in model z4R20. Data are given in units of GeV/nucleon / m$^2$\,s\,sr. On the left the flux for $^{12}$C as a standard primary is shown and on the right the flux for $^{10}$B as a standard secondary.}
\end{figure*}

This effect can be seen by comparing the spatial distribution of $^{12}$C and $^{10}$B as shown in Fig. \ref{FigDist_CarbonBoron_GalPlain}. There the cosmic-ray flux in the Galactic plane for model z4R20 with the Steiman source distribution at an energy of $\sim$10\,GeV/nucleon is shown. As expected, the $^{12}$C flux traces the spiral-arm structure more closely than the $^{10}$B flux. For instance, we find a 40\% decrease of the flux from the nearest spiral arm (the Carina arm) to the position  of Earth for $^{12}$C as compared to 20\% for $^{10}$B.

This also implies that a higher variation of the flux of a primary particle is expected over an orbit of the sun around the Galaxy, which is subject to multiple spiral-arm crossings. Here we find for the $>$ 10\,GeV/nucleon flux of $^{12}$C that $\max{\Phi}/\min{\Phi} \simeq 2$, while for $^{10}$B is is merely $\max{\Phi}/\min{\Phi} \simeq 1.2$. This flux variation for the primaries is consistent with those found, e.g., in the studies
by \citet{EffenbergerEtAl2012AnA547A120, WernerEtAl2015APh64_18} in the context of long-term changes in the cosmic-ray flux at Earth (see, e.g. \cite{Shaviv2003NewA8_39}).

The larger variation of the primary flux along the sun's orbit also implies that secondary-to-primary ratios will show some variation for different positions of the sun along its orbit. This is indeed visible in the different spiral-arm models as will be shown below. Before this variation is discussed,  the simulation results are confronted with the observational constraints.

\subsection{Comparison of numerical results with observational data}
\label{SubSecData}
As was stated in Sec. \ref{SecSetup} all parameter sets used within this study were extracted from \citet{AckermannEtAl2012ApJ750_3}. These parameter sets were tuned to the cosmic-ray fluxes detected at Earth and also to the Galactic diffuse gamma-ray emission. One has to bear in mind that these parameter sets used an axisymmetric source distribution. In the present study this source distribution was simply replaced by one related to the different spiral-arm models, without subsequent alteration of the propagation parameters. This approach was chosen in order to be able to single out the effect of the different source distributions by a comparison of the different models to the results from the model with an axisymmetric source distribution.

\subsubsection{Cosmic-Ray Spectra}

\begin{figure*}
	\setlength{\unitlength}{0.00044\textwidth}
	\begin{picture}(1100,837)(-100,-100)
	\put(360,-70){\small$E_{kin}$/nuc [GeV]}
	\put(-70,-40){\small\rotatebox{90}{$E^2$ flux, (Gev/nuc) m$-^2$\,s$^{-1}$\,sr$^{-1}$}}
	\includegraphics[width=1000\unitlength]{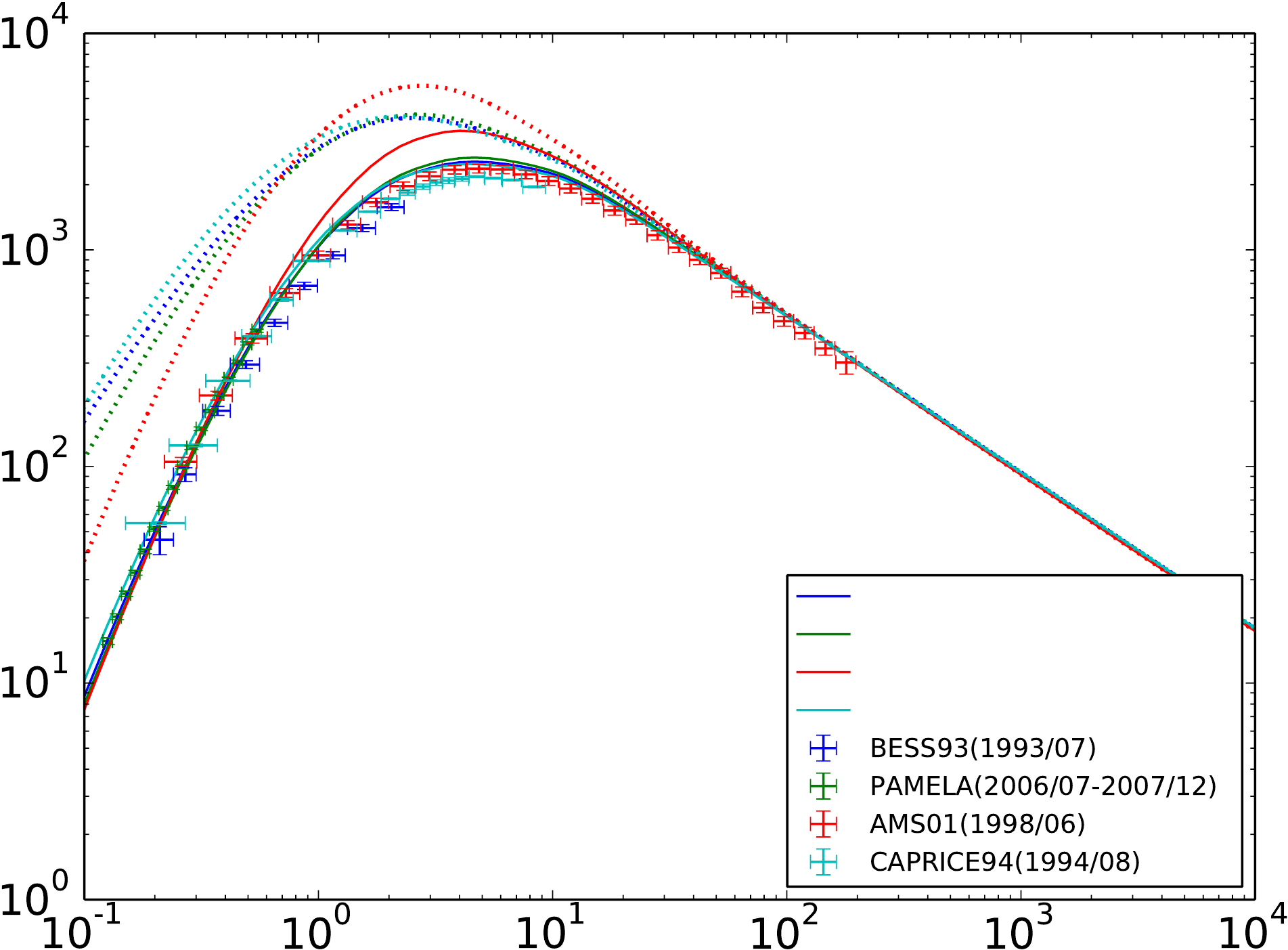}
	\put(-330,264){\tiny Axisymm}
	\put(-330,233){\tiny Steiman}
	\put(-330,203){\tiny Dame}
	\put(-330,173){\tiny NE2001}
	\put(-900,600){\small $^1$H}
	\end{picture}
	\hfill
	\begin{picture}(1100,837)(-100,-100)
	\put(360,-70){\small$E_{kin}$/nuc [GeV]}
	\put(-70,-40){\small\rotatebox{90}{$E^2$ flux, (Gev/nuc) m$-^2$\,s$^{-1}$\,sr$^{-1}$}}
	\includegraphics[width=1000\unitlength]{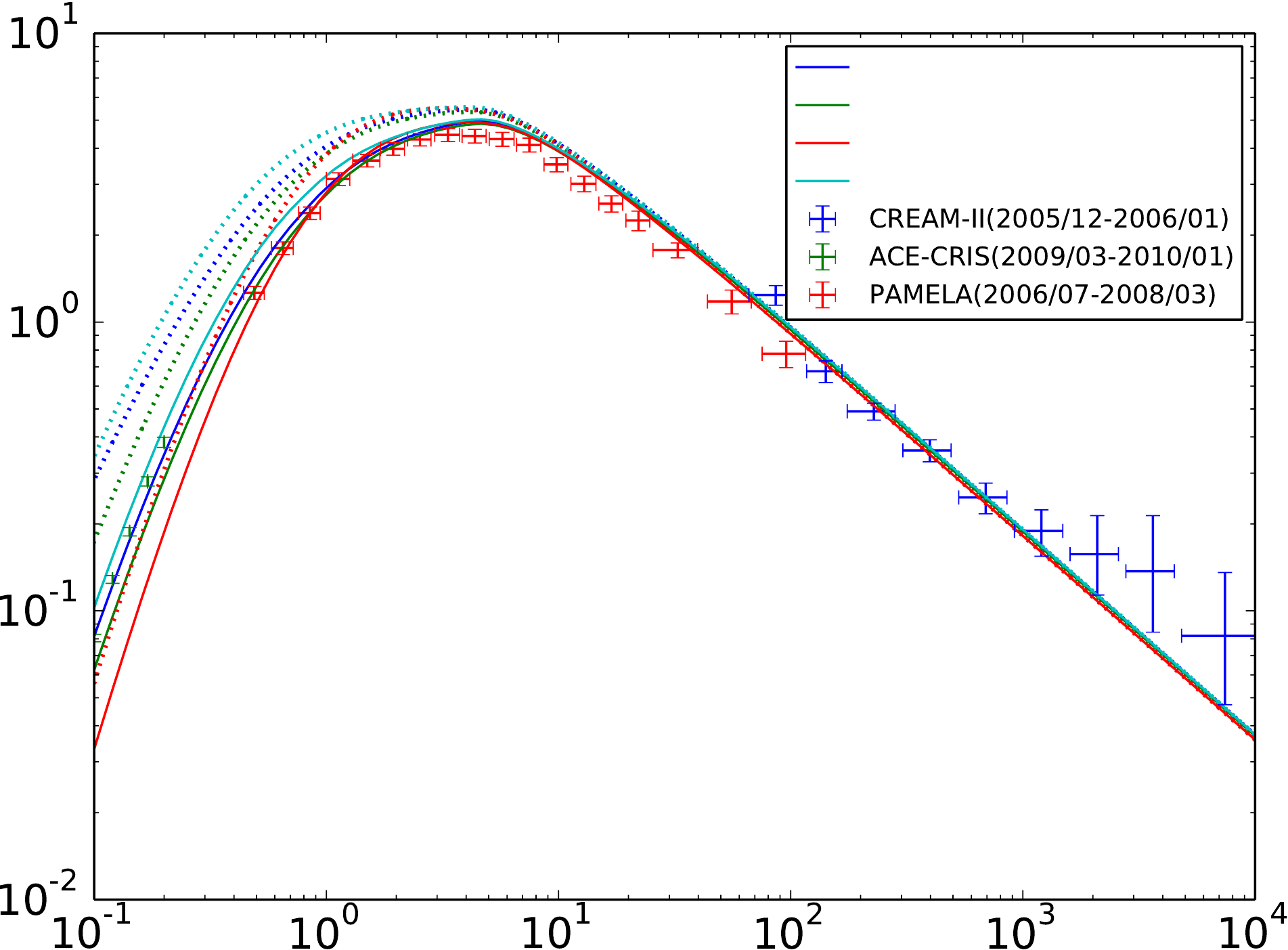}
	\put(-330,679){\tiny Axisymm}
	\put(-330,649){\tiny Steiman}
	\put(-330,619){\tiny Dame}
	\put(-330,589){\tiny NE2001}
	\put(-870,600){\small C}
	\end{picture}
	\caption{\label{FigCRSpecAtEarth}Cosmic-ray flux at the nominal position of the sun for a variety of propagation models. Total protons are shown
on the left and total Carbon
on the right. Results for the z4R20 models are shown using an axisymmetric source distribution (blue), the Steiman four-arm source distribution (green), the Dame two-arm source distribution (red), and the NE2001 source distribution (cyan).
For all models both the unmodulated local interstellar spectrum (LIS, dotted line) and the modulated spectrum (solid line) with a modulation potential $\Phi=450$\,MV for protons and $\Phi=250$\,MV for Carbon is plotted.
	Proton data were taken from \citet{AMS012000PhLB490_27} (AMS-01), \citet{Bess2002ApJ564_244} (BESS), \citet{Boezio1999ApJ518_457} (CAPRICE), and \citet{Pamela2013ApJ770_2} (Pamela).
	 Carbon data were taken from \citet{LaveEtAl2013ApJ770_117} (ACE-CRIS), \citet{AhnEtAl2009ApJ707_593} (CREAM-II) and \citet{Pamela2014ApJ791_93}. }
\end{figure*}
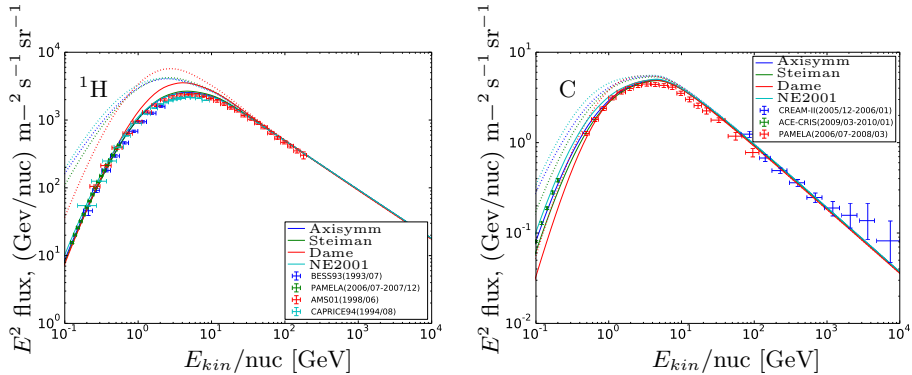

Here we discuss the impact of the different source distribution models by an investigation of the spectra at Earth. As an example, the total proton and the total Carbon spectra are shown in Fig. \ref{FigCRSpecAtEarth}, both as unmodulated LIS and modulated in the force field approximation.
For the latter we use the modulation potential from the respective observational data. Apparently, the spectra for the different source distributions are rather similar with only the Dame source distribution model leading to somewhat larger deviations from the axisymmetric setup. However all model spectra fit  the data just as well as the axisymmetric reference model taken from \citet{AckermannEtAl2012ApJ750_3}.

\begin{figure*}
	\setlength{\unitlength}{0.00044\textwidth}
	\begin{picture}(1100,837)(-100,-100)
	\put(360,-70){\small$E_{kin}$/nuc [GeV]}
	\put(-70,-40){\small\rotatebox{90}{$E^2$ flux, (Gev/nuc) m$-^2$\,s$^{-1}$\,sr$^{-1}$}}
	\includegraphics[width=1000\unitlength]{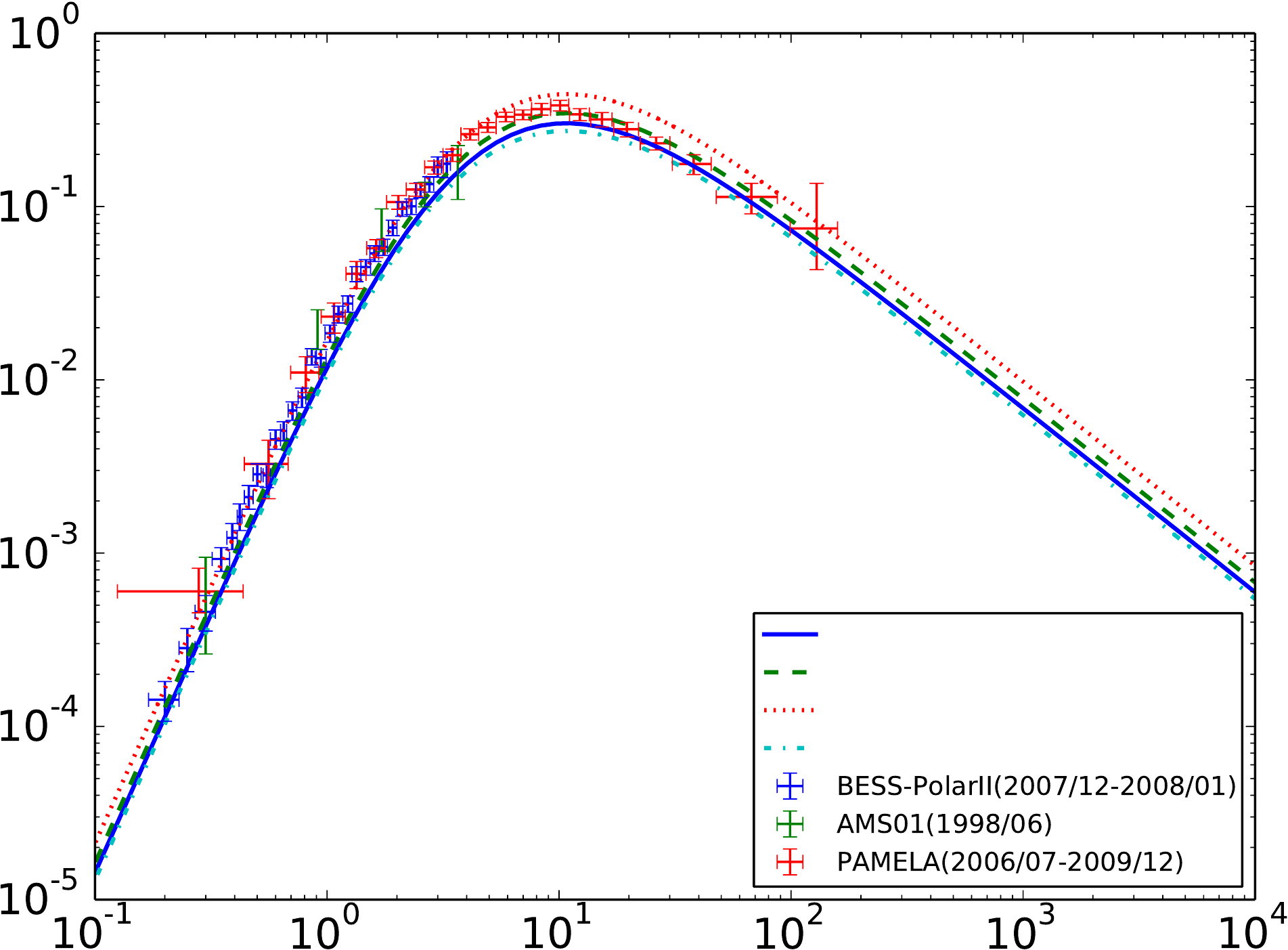}
	\put(-350,227){\tiny Axisymm}
	\put(-350,199){\tiny Steiman}
	\put(-350,171){\tiny Dame}
	\put(-350,143){\tiny NE2001}
	\end{picture}
	\hfill
	\begin{picture}(1100,836)(-100,-100)
	\put(360,-70){\small$E_{kin}$/nuc [GeV]}
	\put(-70,0){\small\rotatebox{90}{flux, m$-^2$\,s$^{-1}$\,sr$^{-1}$\,GeV/nuc}}
	\includegraphics[width=1000\unitlength]{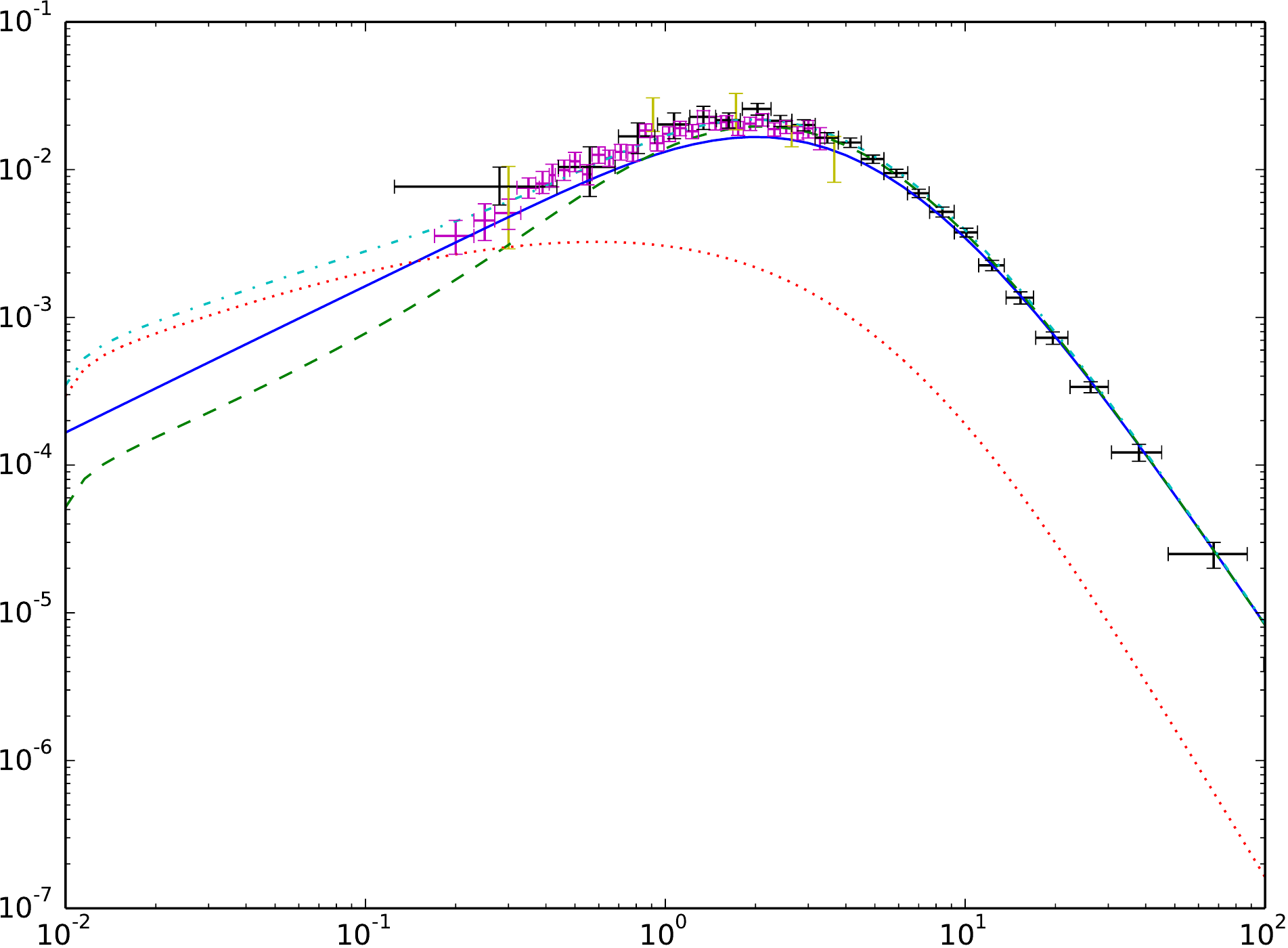}
	\put(-800,550){\small LIS}
	\put(-750,430){\small secondary}
	\put(-360,350){\small tertiary}
	\end{picture}
	\caption{\label{FigAntiprotonSpecEarth}Spectrum of antiprotons at the nominal position of Earth. On the left results for the same four models as in Fig. \ref{FigCRSpecAtEarth} are shown. Corresponding cosmic-ray data were taken from \citet{AMS2002PhR366_331} (AMS01), \citet{BESS2012PhRvL108_1102} (BESS-PolarII) and \citet{Pamela2013JETPL96_621} (Pamela). On the right results for the Steiman z4R20 model only are shown, where additionally the unmodulated data together with the distribution into secondary and tertiary antiprotons is plotted. For the modulated data (solid blue line in the right-handed plot) a modulation potential of $\Phi=350$\,MV was used.
}
\end{figure*}

The antiproton flux shown in Fig. \ref{FigAntiprotonSpecEarth} is considered to be a sensitive cosmic-ray constraint (see, e.g. \cite{MoskalenkoEtAl2002ApJ565_280}). Again, the results of the different models are quite similar. In this case, however the Dame z4R20 model shows a relatively higher antiproton flux at all energies. With the new quality of the data -- in particular those from Pamela (see \cite{Pamela2013JETPL96_621}) -- the results at energies below some 10\,GeV are not in good agreement with the data. Actually, the agreement with the data is fairly good, with $\chi_{dof}^2 = 1.2$ for the axisymmetric model and $\chi_{dof}^2 = 0.8$ for the Steiman z4R20 model, when heliospheric modulation is neglected, pointing to a mismatch at low energies if solar modulation is considered. Application of the force-field approximation with a modulation potential of $\Phi=350$\,MV results in $\chi_{dof}^2 = 6.1$ for the axially symmetric model and $\chi_{dof}^2 = 3.0$ for the Steiman-model. Interestingly, the deviation inherent in our reference model taken from \citet{AckermannEtAl2012ApJ750_3} is somewhat reduced in the Steiman four-arm model. Beyond 10\,GeV both models show a better agreement to the data.

\begin{figure}
\setlength{\unitlength}{0.008cm}
\centering
\begin{picture}(1100,833)(-100,-100)
  \put(390,-70){$E_{kin}$/nuc [GeV]}%
  \put(-70,360){\rotatebox{90}{B/C}}
  \includegraphics[width=1000\unitlength]{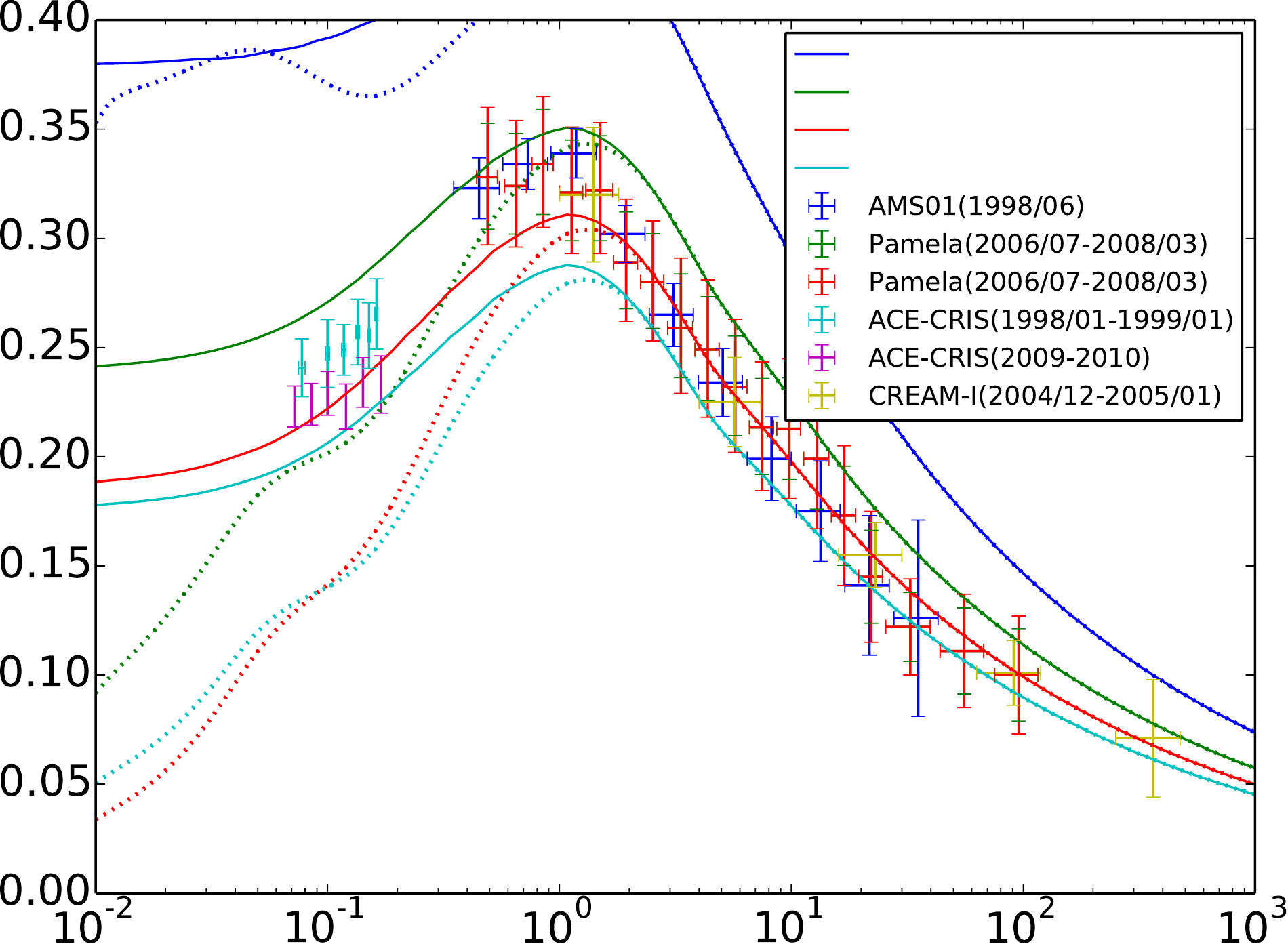}
  \put(-320,683){\footnotesize Dame}
  \put(-320,653){\footnotesize Steiman}
  \put(-320,623){\footnotesize Axisymmetric}
  \put(-320,593){\footnotesize NE2001}
\end{picture}
  \caption{\label{FigBCAtEarth}B/C ratio for four different propagation models. Results for the z4R20 models with the Dame (blue), the Steiman (green), the axisymmetric (red), and the NE2001 (cyan) source distribution models are shown.
For each model the LIS (dotted lines) and the modulated spectrum (solid lines) are plotted, where a modulation potential with $\Phi=350$\,MV was used.
  Data are taken from \citet{deNolfoEtAl2006AdSpR38_1558} and \citet{LaveEtAl2013ApJ770_117} (ACE), \citet{AguilarEtAl2010ApJ724_329} (AMS-1),
     \citet{AhnEtAl2008APh30_133} (CREAM), and \citet{Pamela2014ApJ791_93} (Pamela).
}
\end{figure}
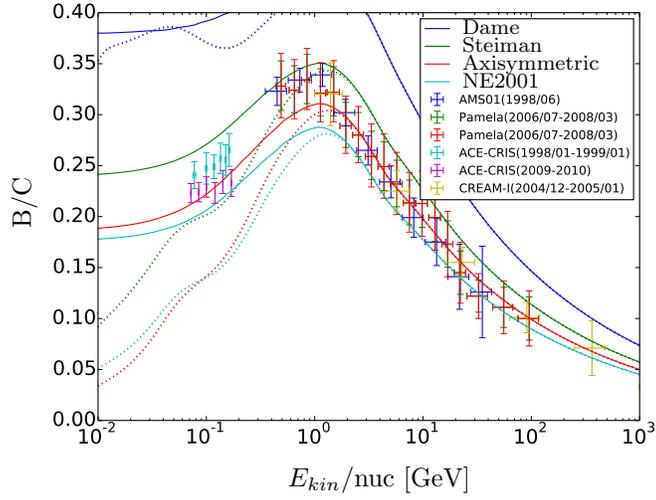

\subsubsection{Secondary To Primary Ratios}
Secondary-to-primary ratios like B/C and $^{10}$Be/$^9$Be constitute an important constraint on cosmic-ray propagation models. While there are other secondary-to-primary ratios, these two are the best measured ones (see, e.g. \cite{StrongEtAl2007ARNPS57_285}). In Fig. \ref{FigBCAtEarth} the B/C ratio for the same models as discussed in previous examples are shown.
The results from the spiral-arm source models do not reproduce the data as well as the results from the axisymmetric model -- 
leading to $\chi_{dof}^2 = 3.57$ for the Steiman-model and $\chi_{dof}^2 = 2.93$ for the NE2001-model as compared to $\chi_{dof}^2 = 0.81$ for the axisymmetric model.
Apparently, the deviation for the Dame-model is much more significant. Again one has to bear in mind that the propagation parameters were tuned to the axisymmetric model, while we note that in particular B/C depends strongly  on the position relative to the spiral arms, as will be discussed below.

\begin{figure}
\setlength{\unitlength}{0.008cm}
\centering
\begin{picture}(1100,831)(-100,-100)
\put(350,-70){$E_{kin}$/nuc [GeV]}%
\put(-70,280){\rotatebox{90}{$^{10}$Be/$^9$Be}}
\includegraphics[width=1000\unitlength]{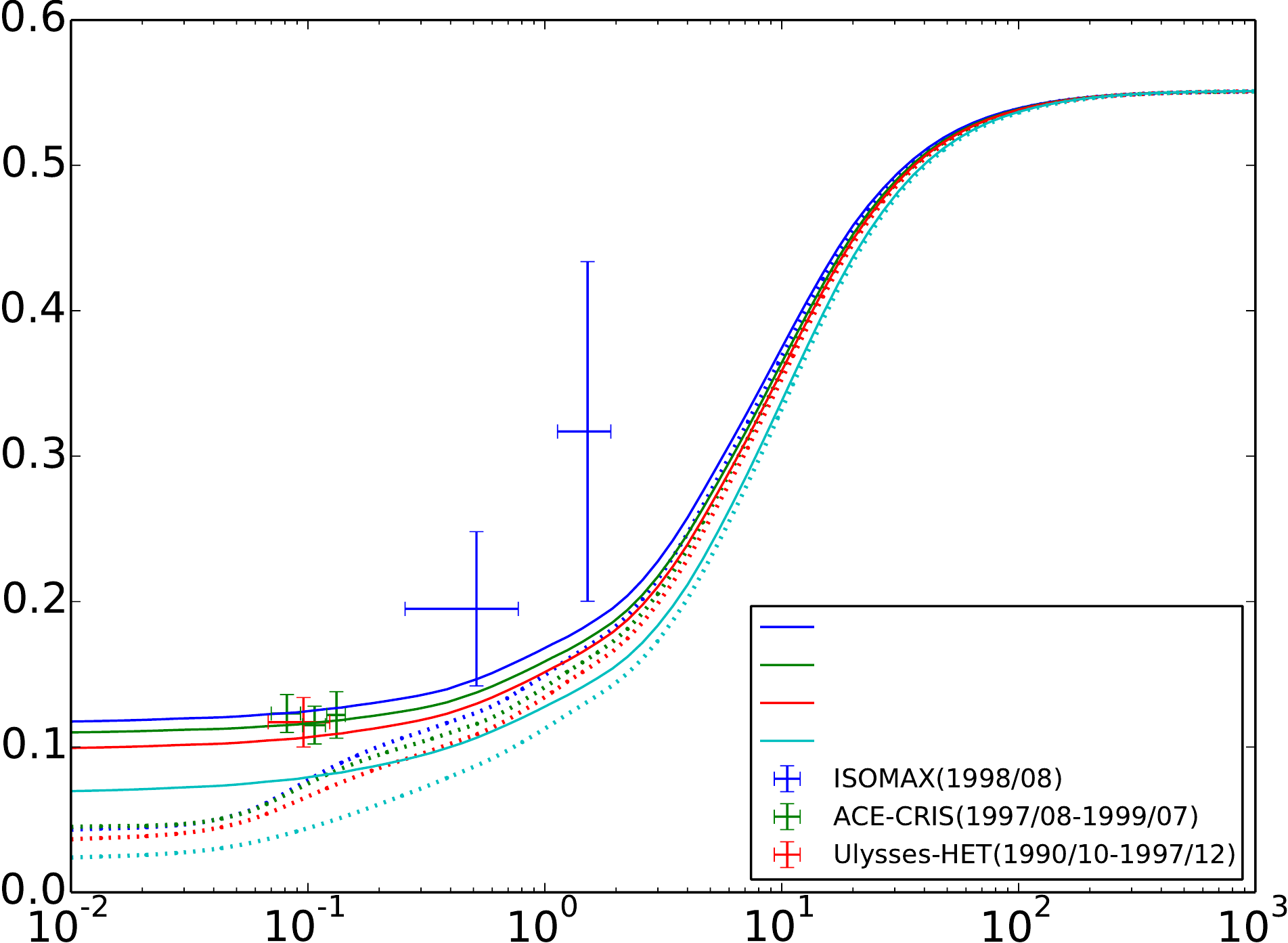}
	\put(-350,231){\footnotesize NE2001}
	\put(-350,201){\footnotesize Axisymmetric}
	\put(-350,171){\footnotesize Steiman}
	\put(-350,141){\footnotesize Dame}
  \end{picture}
  \caption{\label{FigBeRatioAtEarth}Energy dependence of $^{10}$Be/$^9$Be for four different source distribution models (as indicated in the plot) for comparison with corresponding data. These data were taken from \citet{Yanasak2001ApJ563_768} (ACE-CRIS), \citet{HamsEtAl2004ApJ611_892} (ISOMAX) and \citet{Connell1998ApJ501L_59} (Ulysses-HET). The modulated spectra are shown using a modulation potential of $\Phi=550$\,MV, together with the local interstellar spectra (dotted lines).}
\end{figure}

For the $^{10}$Be/$^9$Be ratio shown in Fig. \ref{FigBeRatioAtEarth} the situation is similar to that of the cosmic-ray spectra: the results from the Steiman and the NE2001 source distribution models are very similar to those from the axisymmetric source distribution model, while results from the simulations using the Dame-model show some deviation from the other results. The absolute deviation between the models, however, is much smaller than for B/C. In the next section, the effect of the spiral structure along the sun's orbit will be discussed.

\subsection{Variation of cosmic-ray spectra along Earth's orbit}
Fig. \ref{FigDist_CarbonBoron_GalPlain} illustrates that primary cosmic rays trace the original spiral-arm source distribution more closely than secondary cosmic rays. This also leads to a stronger variation of the primary cosmic-ray flux over sun's orbit around the Galactic centre.
 Here a circular orbit with a radius of 8.5\,kpc at $z=0$\,kpc is assumed.

Along this orbit we find a variation of the $^{12}$C flux above 10\,GeV of a factor of $\max{\Phi} / \min{\Phi} = $2 for the Steiman-model, where the Dame-model gives $\max{\Phi} / \min{\Phi} = $2.3, and the NE2001-model gives $\max{\Phi} / \min{\Phi} = $1.49. For $^{10}$B we find correspondingly $\max{\Phi} / \min{\Phi} = $1.2 for both the Steiman- and the Dame- model and $\max{\Phi} / \min{\Phi} = $1.3 for the NE2001-model. Here, only the Steiman- and the Dame- model show clear indications of spiral-arm crossings, while the orbital variation for the NE2001-model is much smoother \cite{WernerEtAl2015APh64_18}. The relation of the flux observed at the nominal position of sun to the average flux over one orbit is very different in the different models: in the NE2001-model the flux at Earth is some 20\% higher than the average flux, while in the Steiman- and the Dame-model it is 5\% and 15\% lower, respectively. This reflects the presence of the local arm segment in the NE2001-model. Hence only the NE2001-model seems to be consistent with the finding that the cosmic-ray flux in the recent past was some 30\% higher than the long term average (see, e.g. \cite{Shaviv2003NewA8_39}). At this point, however, we have neglected the effect of a time-variable heliospheric modulation and of the influence of the stronger variation of the low energy cosmic rays.

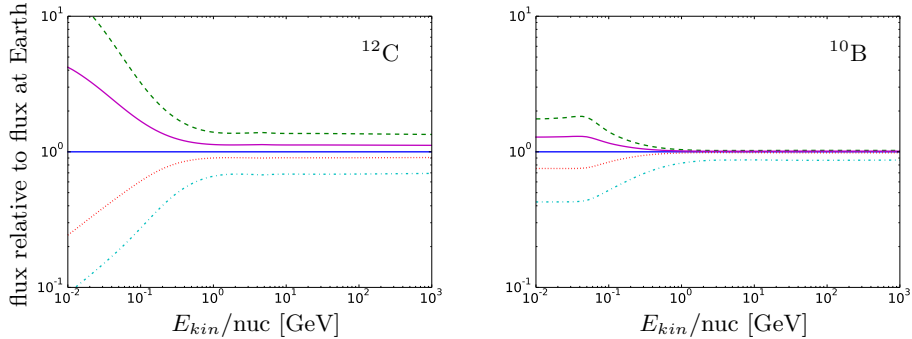
\begin{figure*}
	\setlength{\unitlength}{0.00044\textwidth}
	\begin{picture}(1100,837)(-100,-100)
	\put(330,-70){\small $E_{kin}$/nuc [GeV]}
	\put(-70,-10){\small\rotatebox{90}{flux relative to flux at Earth}}
	\includegraphics[width=1000\unitlength]{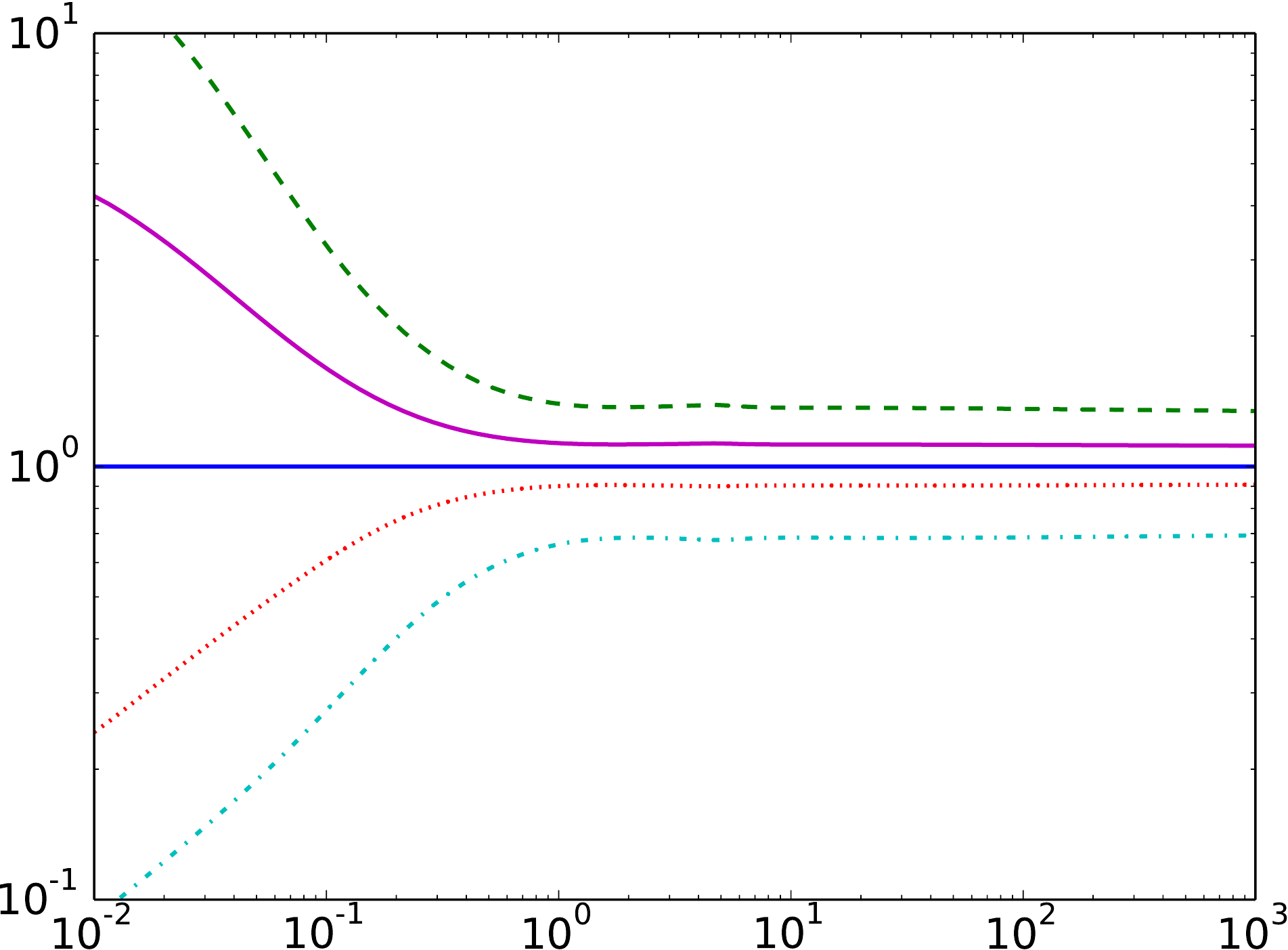}
	\put(-200,600){\small $^{12}$C}
	\end{picture}
	\hfill
	\begin{picture}(1000,837)(0,-100)
	\put(330,-70){\small$E_{kin}$/nuc [GeV]}
	\includegraphics[width=1000\unitlength]{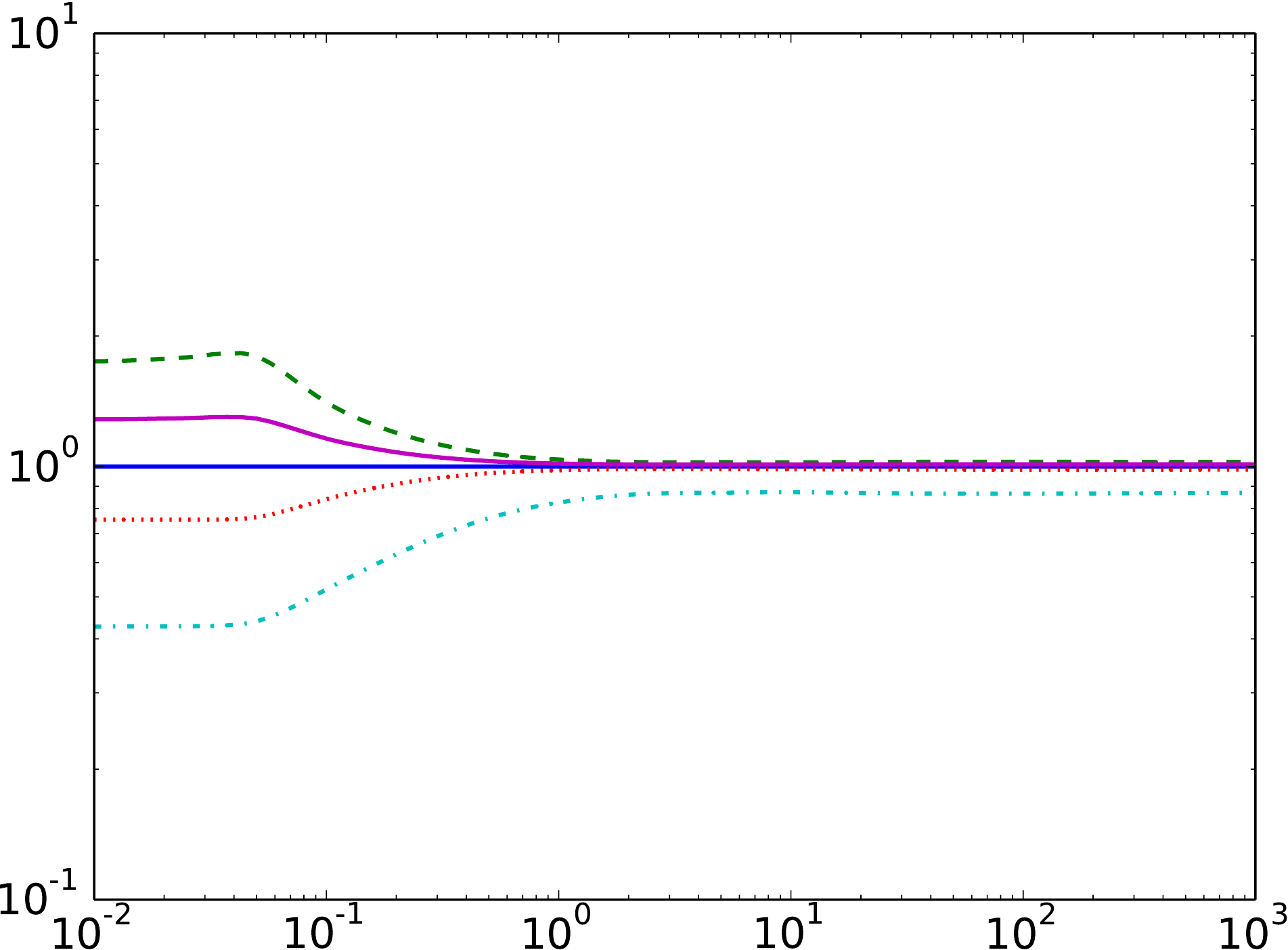}
	\put(-200,600){\small $^{10}$B}
	\end{picture}
	\caption{\label{FigSpecVarPhi}Illustration of the variation of the cosmic-ray spectra along Earth's orbit around the Galactic centre. On the left results for $^{12}$C and on the right such for $^{10}$B are shown.
Deviations relative to the spectrum at the nominal position of Earth (solid blue line) are shown for the nearest spiral arm (green dashed line), the nearest inter-arm position (red dotted line), the widest inter-arm region (cyan dash-dotted line), and an intermediate position in the direction of the Carina arm (solid magenta line).}
\end{figure*}

The variation of the cosmic-ray flux at different energies along the orbit can be visualised by investigating the flux at the positions of the spiral arms and the inter-arm regions. This is done in Fig. \ref{FigSpecVarPhi}, where the flux relative to the flux at the nominal position of Earth for $^{12}$C and $^{10}$B for model z4R20 with the Steiman source distribution is shown at different positions of the orbit.
 The relative flux is shown at the nearest spiral arm ($\phi\simeq 30^{\circ}$), at the nearest inter-arm region ($\phi\simeq340^{\circ}$), at the largest inter-arm region ($\phi\simeq150^{\circ}$) and at an intermediate position in the direction of the nearest spiral arm ($\phi\simeq10^{\circ}$).

As expected, the spectra of the primary cosmic-ray species show a much stronger variation over the spiral arms than for $^{10}$B: for $^{12}$C the variation extends over two orders of magnitude in flux at the lowest energies. The flux is highest at the position of the spiral arms and lowest in the inter-arm region, where the inter-arm region at $\phi=150^{\circ}$ shows the lowest cosmic-ray flux because there the distance to the nearest cosmic-ray sources is largest (see also on the left in Fig. \ref{FigDist_CarbonBoron_GalPlain}). In the Steiman-model the sun is located in the outer regions of the Carina arm. In consequence the flux of the primary cosmic-ray species lies in between the minimum flux at an inter-arm position and the maximum flux at an on-arm position.

\begin{figure*}
  \setlength{\unitlength}{0.00048\textwidth}
  \centering
  \begin{picture}(1000,832)(0,-100)
    \put(320,-70){\small$E_{kin}$/nuc [GeV]}%
    \includegraphics[width=1000\unitlength]{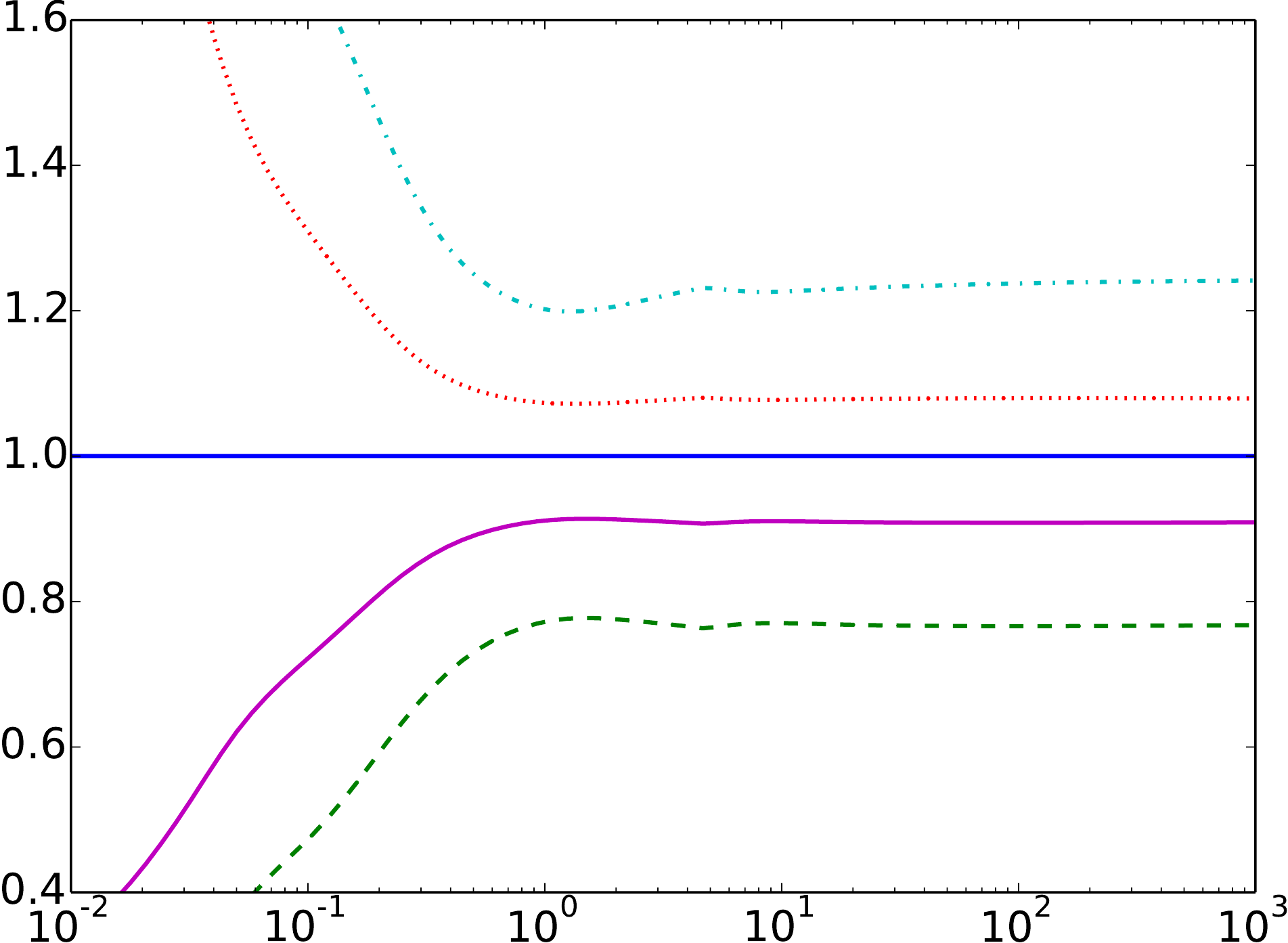}
    \put(-620,600){\small B/C relative to Earth}
  \end{picture}
  \hfill
  \begin{picture}(1000,832)(0,-100)
    \put(320,-70){\small $E_{kin}$/nuc [GeV]}%
    \includegraphics[width=1000\unitlength]{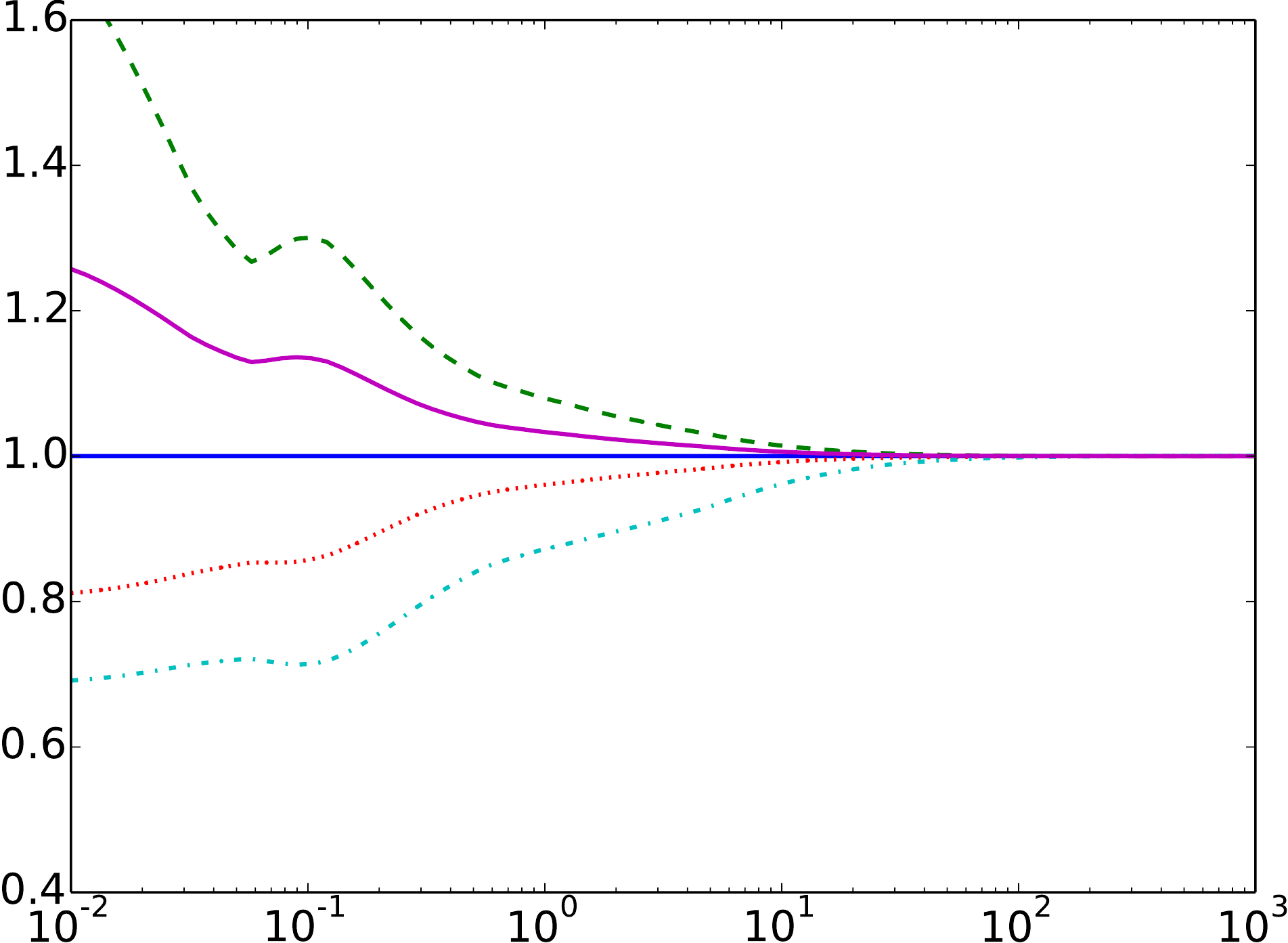}
    \put(-750,600){\small $^{10}$Be/$^9$Be relative to Earth}
   \end{picture}
  \caption{\label{FigRatiosOrbit}
  Secondary-to-primary ratios for the Steiman z4R20 model. Results are shown for the same positions as in Fig. \ref{FigSpecVarPhi}. On the left the energy-dependent deviation of the B/C ratio is shown relative to the value at the nominal position of the sun. On the right is shown the same for the $^{10}$Be/$^9$Be ratio (all unmodulated LIS).}
\end{figure*}

The stronger variation of the primary flux over sun's orbit also leads to a variation of the B/C ratio over this orbit. Corresponding results for the Steiman z4R20 model are shown in Fig. \ref{FigRatiosOrbit}. Evidently there is a strong variation of the B/C ratio over sun's orbit. Both a clear inter-arm and a direct on-arm position results in significant deviations from the B/C ratio at Earth.
The $^{10}$Be/$^9$Be ratio shows much less variation over sun's orbit. This result can be attributed to both species being of secondary nature, only weakly tracing the spiral arms.

\begin{figure*}
\setlength{\unitlength}{0.00045\textwidth}
\centering
\begin{picture}(1100,833)(-100,-100)
	\put(320,-70){\small $E_{kin}$/nuc [GeV]}%
	\put(-70,80){\small \rotatebox{90}{B/C relative to Earth}}
	\includegraphics[width=1000\unitlength]{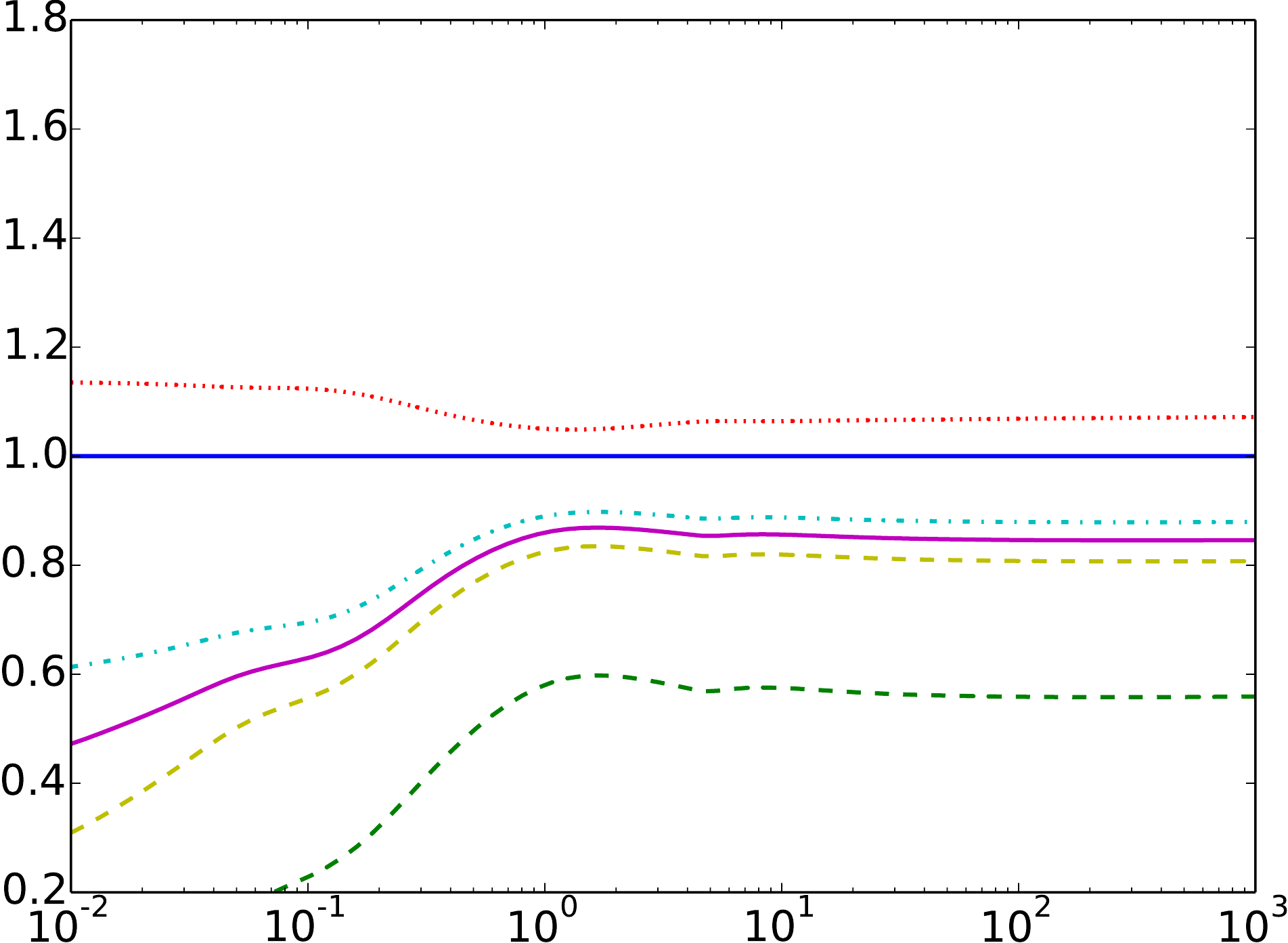}
	\put(-420,600){\small Dame-model}
\end{picture}
\hfill
\begin{picture}(1100,833)(-100,-100)
	\put(320,-70){\small $E_{kin}$/nuc [GeV]}%
	\put(-70,80){\small \rotatebox{90}{B/C relative to Earth}}
  \includegraphics[width=1000\unitlength]{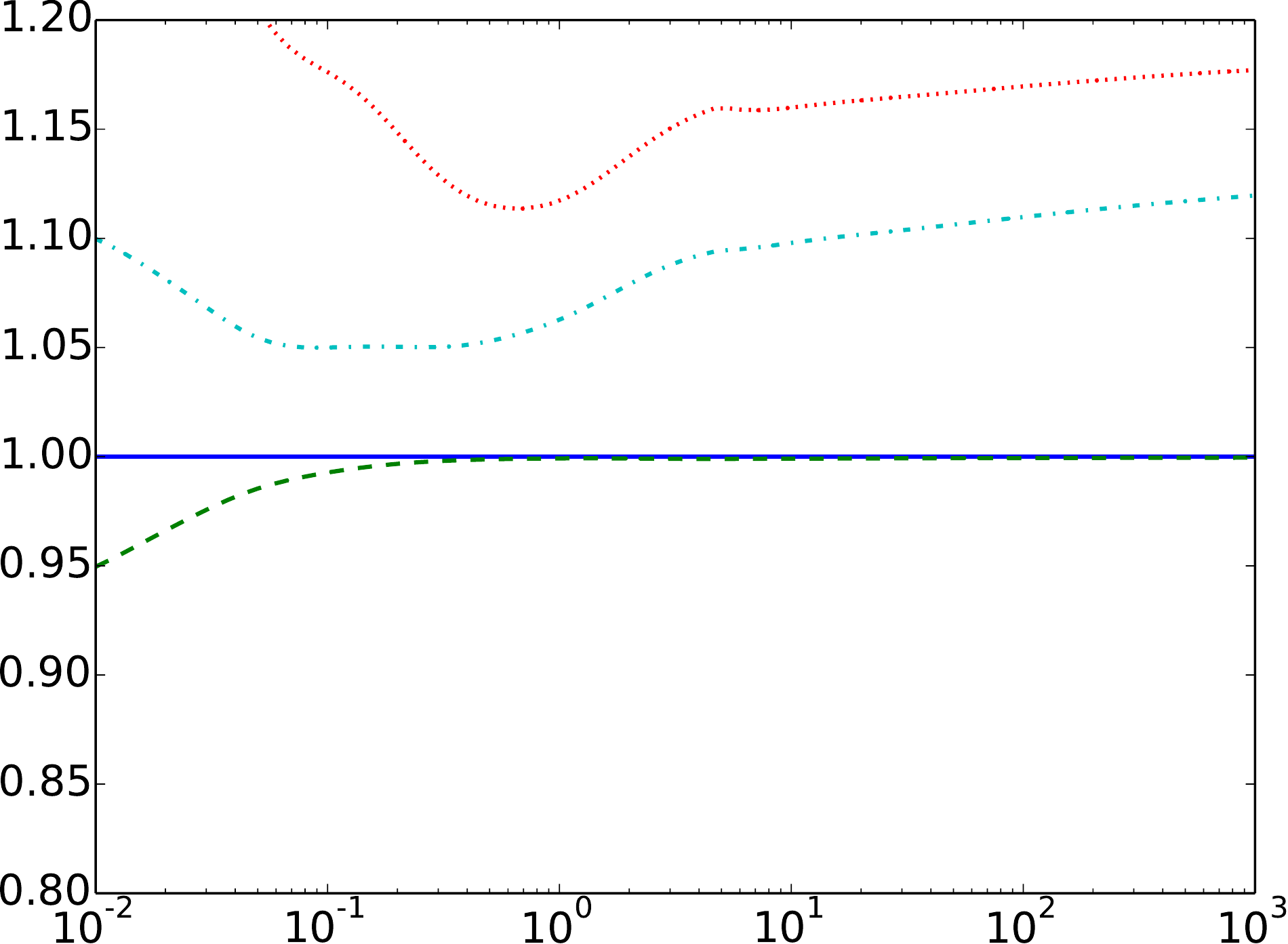}
	\put(-450,450){\small NE2001-model}
\end{picture}
  \caption{\label{FigRatiosOrbitOthers}
  B/C ratios for the Dame two-arm source model (left) and for the NE2001 source model (right) relative to the ratio at the nominal position of the sun. Results are shown for the position of Earth (blue solid line), for an on-arm position (green dashed curve) and for the widest inter-arm position (red dotted line). In case of the Dame-model the remaining results illustrate the variability at the rim of a spiral arm, while the cyan dash-dotted curve shows an intermediate position for the NE2001-model.}
\end{figure*}
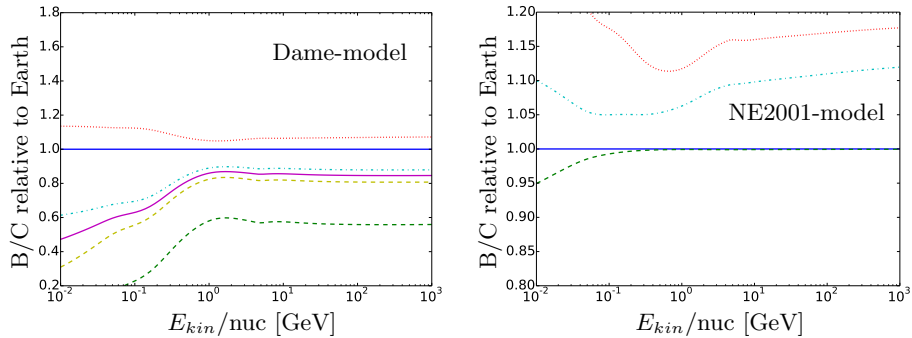

The variation of the B/C ratio along the sun's orbit within the Dame-model is shown in Fig. \ref{FigRatiosOrbitOthers}. Results are given for an on-arm position ($\phi\simeq 90^{\circ}$) at the broadest inter-arm position ($\phi\simeq 180^{\circ}$) and for several positions at the rim of the spiral arm ($\phi\simeq 50^{\circ}, 55^{\circ}, 60^{\circ}$). The latter illustrate the steep local gradient in B/C, where $5^{\circ}$ at a Galactic radius of 8.5\,kpc represent a distance of 742\,pc. The variation along the Earth's orbit becomes more pronounced, because cosmic-ray sources are even more localised in this case: the spiral arms have the same width as in the Steiman-model, but there are only two arms instead of four, leading to much broader inter-arm regions. The most obvious difference to the Steiman-model is, however, that the sun is located in an inter-arm region in this case, thus showing a B/C ratio near the maximum along the sun's orbit.

In contrast, the variation of B/C over sun's orbit is smallest in the NE2001-model (see Fig. \ref{FigRatiosOrbitOthers}). This is caused by the much broader and smoother spiral-arm structures in this model. Here, the presence of the local spiral-arm segment leads to the B/C ratio at Earth being near the long-term minimum.

So far we have ignored the radial epicyclic motion and the vertical motion (see, e.g. \cite{ShavivEtAl2014NatSR4_6150}) of the solar system within the Galaxy.
Of these the vertical motion has potentially the stronger effect, since the sun may pass over instead of through a spiral arm.  We estimate the potential effect by comparing the flux inside and 125\,pc above a spiral arm. This leads to the conclusion that the effect can be ignored in this context, since, e.g. for B/C it is one order of magnitude smaller than the contrast between the nominal position of the sun and the nearest spiral arm.

\subsection{Consequences of source localisation}
For the standard set of propagation parameters the B/C ratio, and in case of the Dame-model also the $^{10}$Be/$^9$Be ratio, at the nominal position of sun is inconsistent with the data for all spiral-arm source models. In the previous section it was found that not only the fluxes but also these ratios vary in space to a certain degree. This leads to the question whether the gradients in these quantities are sufficiently large in the immediate neighbourhood of sun to allow a fit to the data within the uncertainties of the spiral-arm models (see the discussion in Sec. \ref{SecSetup}). Possible uncertainties include the location of the relevant tracer within the spiral arm, the shift due to the delay between spiral-arm passage and occurrence of the supernova explosions, and also the distance of the sun from the Galactic centre. The latter would result in a rescaling of the entire Galactic structure.

We test this by introducing an azimuthal shift of the spiral-arm structure,  to emulate the delay of the supernova explosions. This test shows very different results for the different source models. For the Steiman-model the azimuthal gradient in the vicinity of the sun is so large that a shift of the spiral-arm pattern by just $\Delta \phi=-10^{\circ}$ leads to an excellent fit for B/C with $\chi_{dof}^2 = 0.57$. The related change in B/C was found to be near 10\% at high energies and near 20\% at 10\,MeV. This rotation of the spiral-arm pattern brings cosmic-ray sources from the Carina arm closer to sun, which is consistent with the offset between supernovae and the peak in the gas density within the spiral arms (see \cite{Shaviv2003NewA8_39}).

For both the Dame- and the NE2001-model a satisfactory fit in the vicinity of the sun cannot be found: for both models the gradient in B/C near Earth is rather shallow. Tests showed that a fit can only be found by extreme shifts of the spiral-arm pattern in the azimuthal direction that are not within possible uncertainties for the source distribution. In all cases, however, a good fit is found when the rim of a spiral arm coincides with the location of the sun. The discrepancy very likely relates to the location of sun in a spiral arm in the NE2001-model and far away from the nearest arm in the Dame-model.

Because of the small pitch angle of the spiral arms, typically around 12$^{\circ}$, the variation of the cosmic-ray flux in the radial direction is typically higher in these models. For the Steiman source distribution model a shift of 500\,pc in the direction of the Galactic centre leads to a change of 15\% in B/C even at highest energies.
Thus the observed B/C ratio in this case is only representative for the immediate vicinity of the sun. Hence local observations do not give  good constraints for a Galactic propagation model.
For the other source distribution models, however, the gradient is much smaller, amounting to a change of B/C at high energies of some 3\% for both the NE2001- and the Dame-models. In these cases the observed B/C ratio is representative for a larger region of the Galaxy. Especially for the Dame-model this is only true within the inter-arm regions. Near the cosmic-ray sources the gradient in B/C is of the same order or even higher than for the Steiman-model.

This discussion strengthens the conclusion that the B/C ratio, as the most commonly referred secondary-to-primary ratio, is subject to strong spatial variation depending on the distance from the cosmic-ray sources. Depending on the model, secondary-to-primary ratios are sometimes only a local measure and  may or may not be applicable throughout the Galaxy. Furthermore, this finding shows that more detailed propagation models will also require modfication of some of the other propagation parameters. 
Therefore very low resolution parameter studies do not necessarily translate into more realism in the model.
In contrast, the $^{10}$Be/$^9$Be ratio, being a ratio of two secondaries,  was found to be a more global measure and hence more useful in constraining cosmic-ray transport parameters.
   This statement does not extent to all cosmic-ray clocks. As an example, $^{26}$Al/$^{27}$Al shows significant variation in the 10\,GeV range, since only $^{27}$Al is a primary in this case.

\begin{figure*}
\setlength{\unitlength}{0.0006\textwidth}
\centering
\begin{picture}(1100,837)(-100,-100)
	\put(360,-70){\small $E_{kin}$/nuc [GeV]}%
	\put(-70,280){\small \rotatebox{90}{Anisotropy}}
	\includegraphics[width=1000\unitlength]{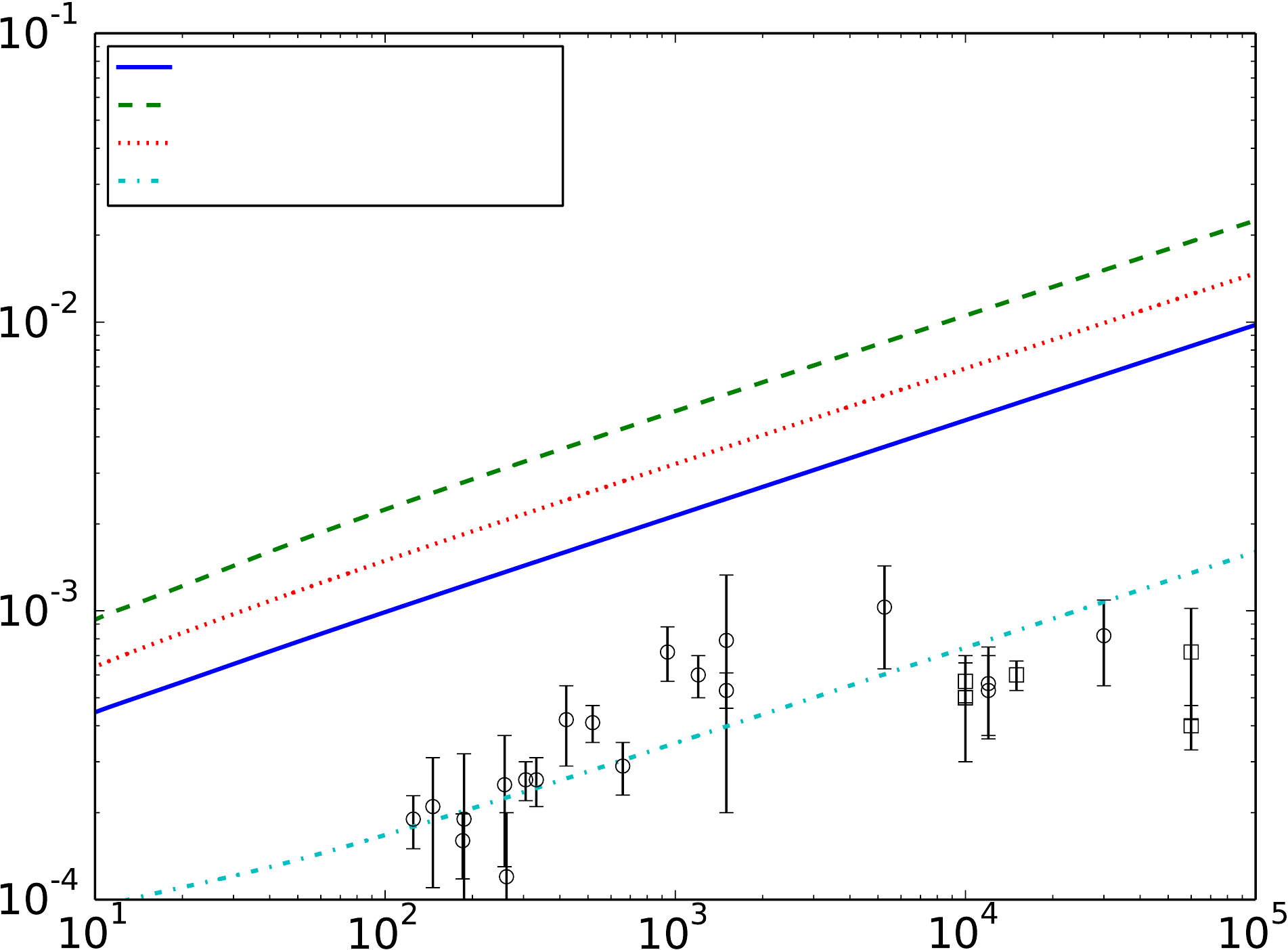}
	\put(-860,678){\tiny Axisymmetric}
	\put(-860,648){\tiny Steiman}
	\put(-860,618){\tiny Dame}
	\put(-860,588){\tiny NE2001}
\end{picture}
  \caption{\label{FigCRAniso}
	Dipole amplitude of the cosmic-ray anisotropy as computed for the different source distribution models. Here we show results for the axisymmetric source distribution model (blue solid line), the Steiman-model (green dashed line), the Dame-model (red dotted line), and the NE2001-model (cyan dash-dotted line). Data are taken from \citet{GuillianEtAl2007PhRvD75_2003} -- see also references therein. Squares indicate data from  extensive air-shower arrays and circles indicate data from muon detectors.
}
\end{figure*}
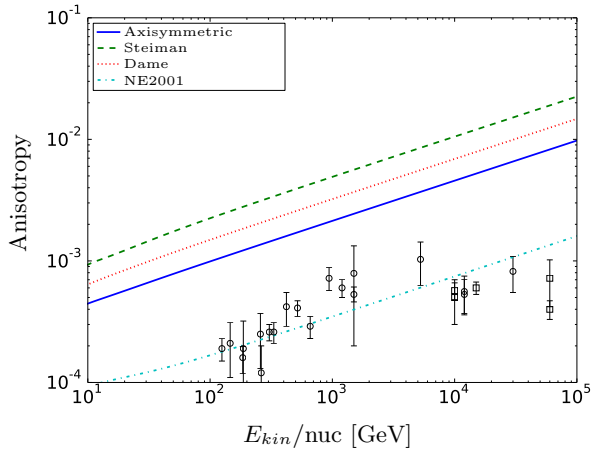   

The range in different gradients for the cosmic-ray flux so far encountered motivates an investigation of the resulting cosmic-ray anisotropy. In Fig. \ref{FigCRAniso} we show the computed dipole amplitude of the cosmic-ray anisotropy in comparison to observations. Results for the axisymmetric model illustrate a common problem in Galactic cosmic-ray propagation models (see, e.g. \cite{EvoliEtAl2012PhRvL108_1102}): the anisotropy in most propagation models clearly exceeds the observed anisotropy in the $\sim$1\,TeV range. The problem is even more severe for the Steiman-model due to the vicinity of the cosmic-ray sources in the Carina arm. While results for the Dame-model show a somewhat smaller anisotropy due to the large distance to the next spiral arm, results for the NE2001-model interestingly are in good agreement with the data. This is caused by the presence of the local spiral-arm segment, reducing the cosmic-ray gradient towards TeV energies. This is another strong hint that the local spiral-arm segment should be considered in models for the Galactic cosmic-ray source distribution.

\section{Constraints on propagation parameters}
We have shown how different source distribution models fit the cosmic-ray data without changing the principal propagation parameters. Whereas the Steinman-model is in good agreement with the data, when taking into account the uncertainties in the spiral-arm models, the NE2001-model indicates that the local spiral-arm segment is important in explaining other observables, and the Dame-model is inconsistent with the data when using the reference propagation parameter set. Now we investigate and discuss the impact of individual propagation parameters on the models.

\subsection{Influence of the halo height}
The first parameter under consideration is the height of our Galactic halo, typically estimated with values between 4-6 kpc but recently challenged with  values up to 10 kpc (see \cite{OrlandoStrong2013MNRAS436_2127}). To investigate the influence of different halo heights,
 results from two Steiman-models with identical radial extent but different halo heights were investigated. This allows for a direct comparison of the results. Following  the discussion in \ref{AppendResStudy} model z8R30 is used for the investigation of the effects of a larger halo height, i.e. using the model with the larger radial extent.
To avoid any influence of the radial boundary, this model will be compared to model z4R30; the number of grid points was doubled in model z8R30 in the $z$-direction,  giving the same spatial resolution for both models.

\begin{figure*}
  \setlength{\unitlength}{0.00072\textwidth}
  \centering
  \begin{picture}(1100,832)(-100,-100)
    \put(390,-70){\small$E_{kin}$/nuc [GeV]}%
    \includegraphics[width=1000\unitlength]{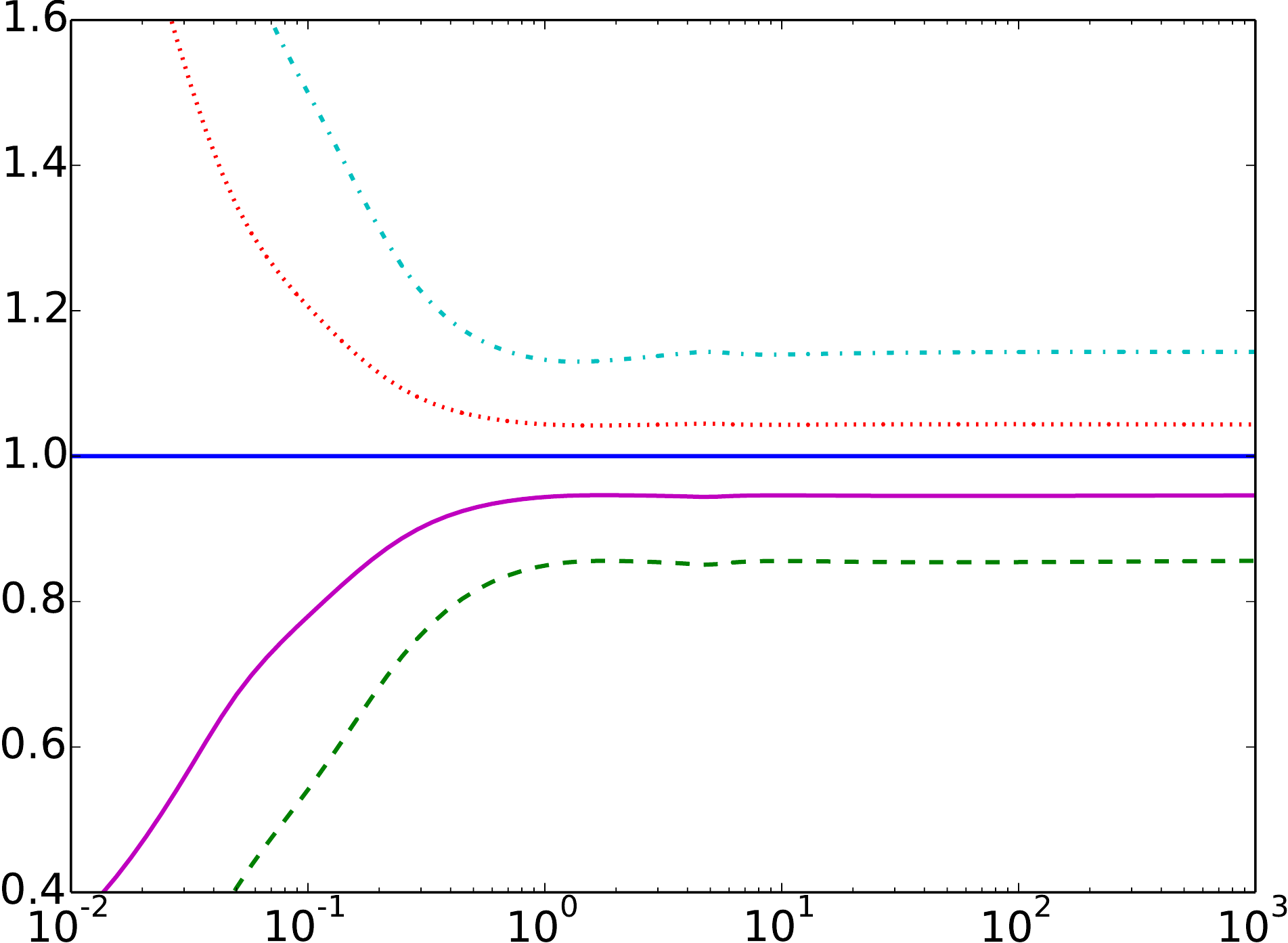}
    \put(-450,650){B/C relative to Earth}
  \end{picture}
  \caption{\label{FigRatiosOrbitHalo}
  Same as left part of Fig. \ref{FigRatiosOrbit}, now for the
  for the Steiman z8R30 model.}
\end{figure*}

Fig. \ref{FigRatiosOrbitHalo} shows the variation of the B/C ratio along sun's orbit for model z8R30. Results for model z4R30 were found to be indistinguishable from those from model z4R20 shown in Fig. \ref{FigRatiosOrbit}
A comparison of Figs. \ref{FigRatiosOrbit} and \ref{FigRatiosOrbitHalo} shows that a larger halo height decreases the variation of B/C along the sun's orbit, i.e. we find a variation of more than a factor 1.5 for $z=4$\,kpc and less than 1.35 for $z=8$\,kpc. However the same qualitative behaviour as for the models with the smaller halo height is observed: again, a fit to the data is found by a shift of the spiral-arm pattern with $\Delta \Phi=-10^{\circ}$, i.e., when the sun is located at the rim of the source region related to the Carina arm.

This particular difference between large and small halo height models can be easily understood when investigating the model parameters given in Tab. \ref{TabParameters}. Due to the degeneracy between diffusion coefficient and halo height (see also the discussion in \cite{TrottaEtAl2011ApJ729_106}), the diffusion coefficient needs to be higher in  models with a larger halo height. Such a larger diffusion coefficient at the same time implies a stronger broadening of any spatial structures. This means that the model results show smoother spiral arms in the cosmic-ray distribution both in the Galactic plane and also in the vertical direction.
This is also reflected in the variation of the interstellar cosmic-ray flux over the solar orbit around the Galactic centre. Where for the $z=4$\,kpc model we find for the variation of the $^{12}$C flux above 10\,GeV that $\max {\Phi}/\min{\Phi}$=2, this value is reduced to 1.47 for the $z=8$\,kpc model, i.e., the amplitude of the flux variation is reduced by a factor of 2. 
This is consistent with the stronger orbital variation in models with a smaller halo height \cite{Shaviv2003NewA8_39}.

\subsection{Alternative Propagation Parameters}
Comparison of the results of models with different halo heights shows that variation of B/C over the sun's orbit decreases with increasing  spatial diffusion coefficient. The spectrum at low energies is determined from an equilibrium of spatial diffusion, diffusive re-acceleration and energy-loss processes (see, e.g. \cite{StrongEtAl2007ARNPS57_285}). Since the latter are thought to be well understood, changing one or both of the former two should help to find a better fit to the data.

As shown in the foregoing discussion, with the reference parameter set the B/C ratio at an on-arm position is too low, while being too high at an inter-arm position. A satisfactory fit to the data is only found when the spiral arms are oriented in a way that Earth is located at the rim of the spiral arm, i.e., at a position where the flux has decreased somewhat from the high level observed at an on-arm position. In the four arm model B/C at the nominal position of Earth is still somewhat too high. Thus a fit could be found by shifting the spiral-arm pattern in a way that sun is located nearer to the spiral arms.

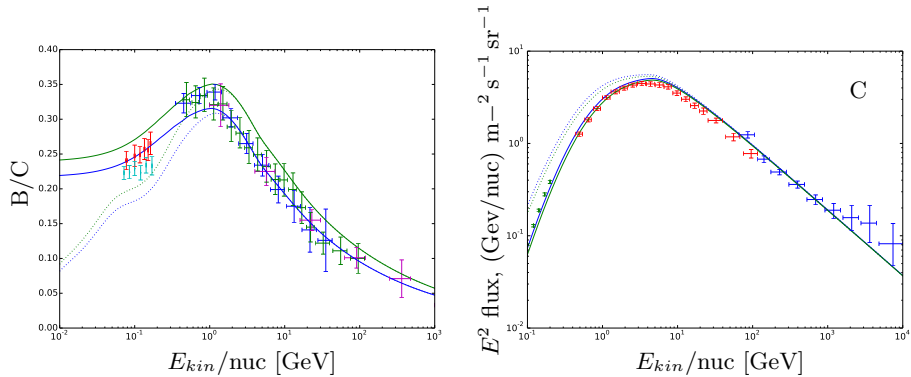
\begin{figure*}
  \setlength{\unitlength}{0.00045\textwidth}
  \centering
  \begin{picture}(1100,832)(-100,-100)
    \put(320,-70){\small$E_{kin}$/nuc [GeV]}%
    \put(-70,330){\small\rotatebox{90}{B/C}}
    \includegraphics[width=1000\unitlength]{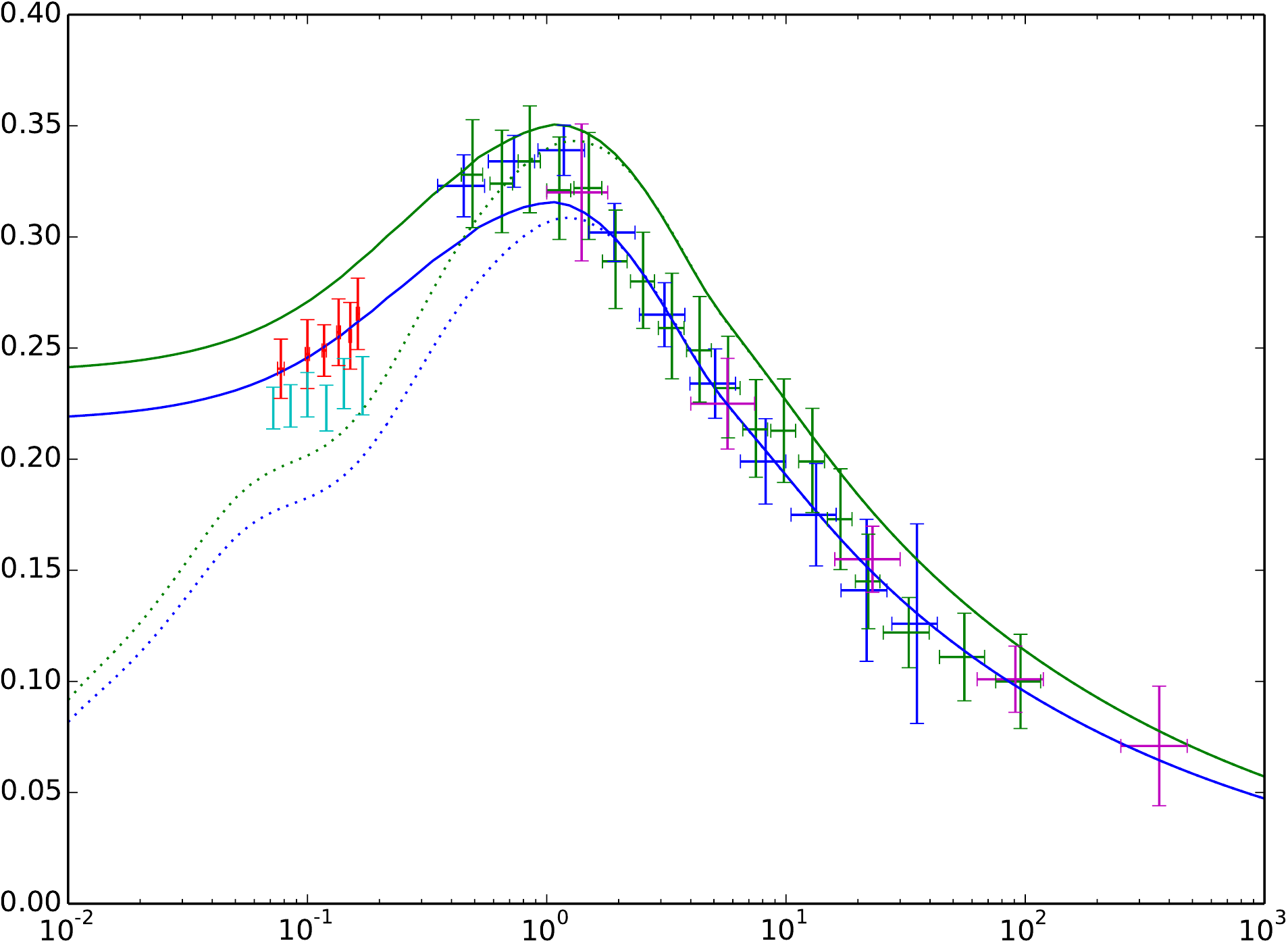}
  \end{picture}
  \hfill
  \begin{picture}(1100,835)(-100,-100)
    \put(320,-70){\small $E_{kin}$/nuc [GeV]}
    \put(-70,-40){\small\rotatebox{90}{$E^2$ flux, (Gev/nuc) m$-^2$\,s$^{-1}$\,sr$^{-1}$}
    }
    \includegraphics[width=1000\unitlength]{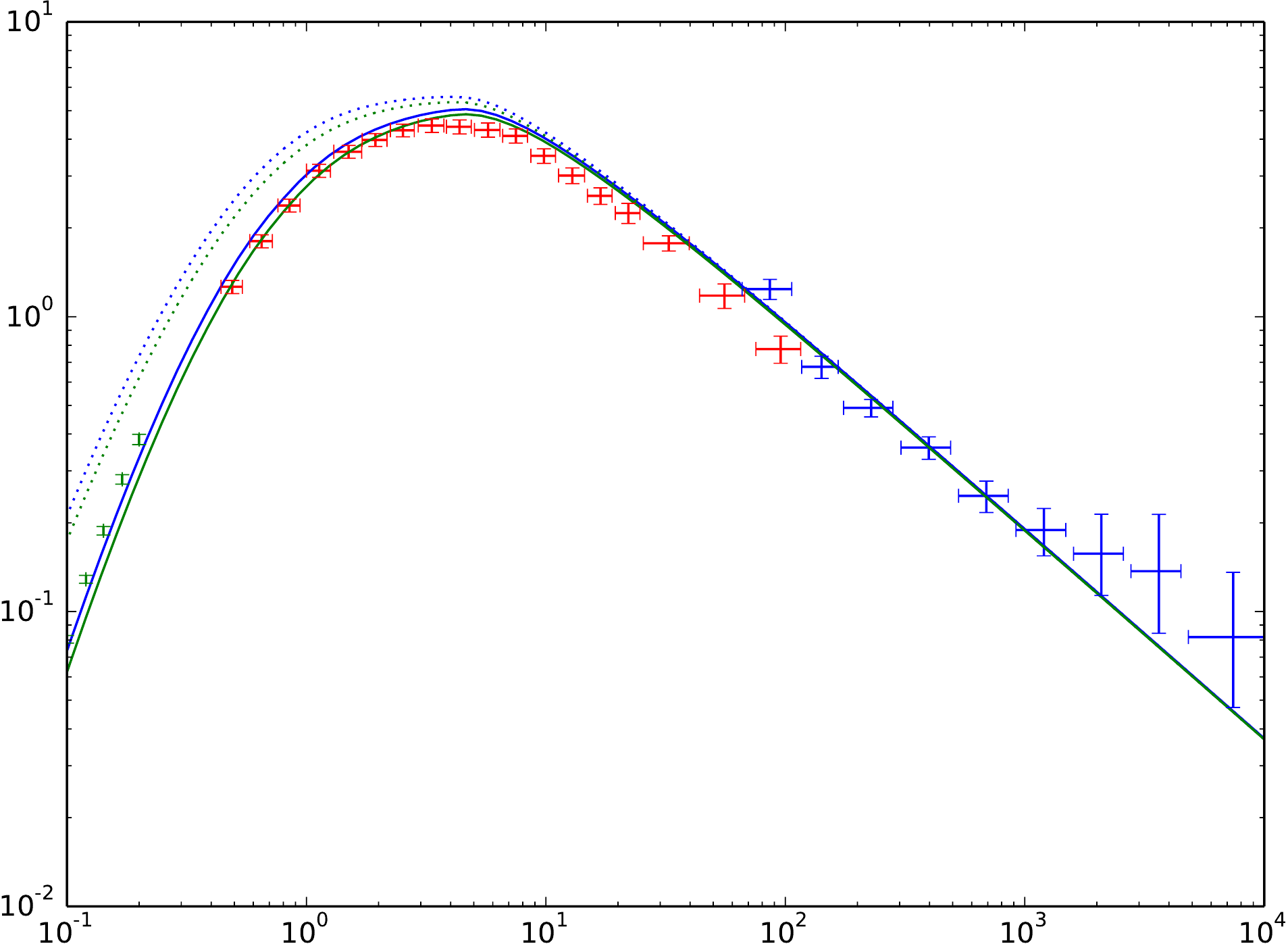}
    \put(-150,600){\small C}
  \end{picture}
  \caption{\label{FigVarModelEarth}
  Effects of a change in the transport parameters. Results from a standard z4R20 Steiman-model (green) are compared with a model where a modified set of propagation parameters was used (blue). On the left the energy dependence of B/C (modulation potential $V=350$\,MV) is shown and on the right the total Carbon spectrum (modulation potential $\Phi=250$\,MV). Results are given for the nominal position of the sun.}
\end{figure*}

To find a fit to the observed cosmic-ray data without invoking a rotation of the spiral-arm pattern, we changed the diffusion coefficient $D_0$ and the strength of the re-acceleration, i.e. the magnitude of $v_A$. Fig. \ref{FigVarModelEarth} shows results for model z4R20 with the Steiman
 source distribution with $D_0=6.5$ and $v_A$=38\,km\,s$^{-1}$,  compared to
results for the standard propagation parameters. The agreement of the model  using this modified parameter set with observations for B/C has improved leading to $\chi_{dof}^2 = 0.72$, while cosmic-ray spectra show almost no difference, as can be seen from Fig. \ref{FigVarModelEarth} for total Carbon. This is due to the rather small change in the propagation parameters, which also keeps $^{10}$Be/$^9$Be nearly unchanged.

Using the NE2001-model, the propagation parameters need to be changed in the opposite direction since the sun is located on a spiral-arm segment instead of in an inter-arm regions. In this case a fit of similar quality ($\chi_{dof}^2 = 0.63$) is found by using $D_0=4.2$ and $v_A=26$\,km\,s$^{-1}$. We note that this results in a slightly further reduction of the anisotropy.

For the Dame-model we did not find a satisfactory fit. While the fit to the data at high energies was good, when using $D_0 = 10$ and $v_A = $58.6\,km\,s$^{-1}$, neither the peak around 1\,GeV nor the low-energy behaviour was captured by the results. Apart from that the necessary parameters, especially the Alfv\'en speed appear rather extreme. This is a further indication that the Dame-model is not consistent with cosmic-ray data. But taking the local spiral-arm segment into account in the Dame-model has the potential to qualify such a drastic conclusion.

Apparently a set of best-fit propagation parameters can be found for models with different spiral-arm source distributions, while a model with very distant sources  seems to be excluded.
This shows that the propagation parameters in a model with a more local cosmic-ray source distribution are different from such with an axisymmetric source distribution. Thus, propagation models with a higher complexity in the incorporated propagation physics can be expected to lead to different values for the propagation parameters.

It appears to be possible to find comparably good fits to the cosmic-ray data at the location of sun for models with rather different source distribution, which also lead to different spatial distributions of the cosmic-ray flux. Therefore the quality of a propagation parameter set needs to be assessed through a measure of the spatial distribution of the cosmic-ray flux as well. This is provided by the Galactic diffuse gamma-ray emission, and the search for a best-fit propagation parameter set will need to consider this emission alongside the cosmic-ray data.

\section{Conclusion}
In this study we have investigated the effects of the nuclear reaction network in the context of spiral-arm cosmic-ray source distributions. To allow a study of comparably small-scale structures a high-resolution three-dimensional propagation code was applied, the recently introduced \textsc{Picard} code.
 Initially, all models were computed using a set of established propagation parameters. For all spiral-arm models we found that the deviations for the resulting cosmic-ray fluxes at Earth are rather small when compared to axisymmetric models. In the case of the 2-arm model such deviations become strikingly evident.

A more sensitive tracer of the differences resulting from the non-axisymmetric source distribution are the secondary-to-primary ratios. We concentrated on the B/C ratio as being one of the best-measured of these. At the same time it turns out to be one of the most sensitive of these ratios to spatial variations of the cosmic-ray source strength. In B/C only the Steiman four arm source distribution model fits to the data at a similar, albeit somewhat inferior, level compared to the axisymmetric model. Especially the two-arm model shows large deviations from the data at the nominal position of the sun.

Assuming that the chosen set of transport parameters is correct, we found that a satisfactory fit can be found in any of the spiral-arm source distribution models, when the spiral arms are oriented in a way that sun is located at the rim of the nearest spiral arm. Both an inter-arm and an on-arm position lead to inconsistent results. 
Such a setup can only be reconciled within the model uncertainties of the Steiman-model.

This study shows that
the B/C ratio is a rather local quantity, which can be subject to significant variations even on small spatial scales. 
For models with strong gradients in the source distribution near the sun,  secondary-to-primary ratios only reflect the cosmic-ray distribution in the immediate vicinity of the sun.
Unfortunately the actual gradients in the cosmic-ray source distribution are not precisely known.
Ratios of secondaries or ratios of primaries show less spatial variation even in the presence of a spiral-arm source distribution.
 This also shows that the position of the sun is not necessarily representative with regard to the cosmic-ray flux. With propagation parameters without spatial variation, an inhomogeneous source distribution will results in a significant spatial variation in the B/C ratio. This finding has to be seen in the context that recent models can only reproduce the electron and positron data when taking local cosmic-ray sources into account (see, e.g. \cite{GaggeroEtAl2013PhRvL111_021102}).
 
We also investigated the possibility to find alternative propagation parameter sets for different spiral-arm source distributions, by allowing alterations of the strength of spatial diffusion and re-acceleration. We found acceptable parameter sets for the Steiman- and the NE2001-model, but there is  tension between the Dame-model and the cosmic-ray data.
Comparison of fits to cosmic-ray spectra cannot presently distinguish between the Steiman- and the NE2001-model.
We note that the NE2001-model can recover the observed cosmic-ray anisotropy in the 1\,TeV range, whereas the Steiman-model shows a recurrent imprint of the spiral-arm pattern on the cosmic-ray flux over the sun's orbit around the Galaxy. This would connect to the findings by \citet{Shaviv2003NewA8_39} that such a variation is visible in meteorite records. On the other hand these data also indicate that the current cosmic-ray flux represents a long-term maximum, which is only recovered in the NE2001-model. We find that only a model combining the different features of the investigated spiral-arm models might be able to explain all available data. Such a model might combine the spiral-arm structure of the Steiman-model with an additional local spiral-arm segment.

In this study we investigated cases where cosmic-ray sources were either distributed axisymmetrically or confined to spiral arms.
Thus, it might seem that such results correspond to our choice of extreme cases for the arrangement of cosmic-ray sources and the variation along sun's orbit might be weaker in reality. 
However we  stress that it will also be worthwhile to investigate the effect of other alternative parameters in cosmic-ray propagation models. These include the Galactic gas distribution (see, e.g. \cite{JohannessonEtAl2013ICRC}) or the structure in the interstellar magnetic field (see, e.g. \cite{OrlandoStrong2013MNRAS436_2127, Jansson2012ApJ757_14}). These will also have effects on cosmic-ray particle propagation and also be considered in the future.

To conclude, we will advance into more detailed propagation models, where \emph{more detailed} refers to an inclusion of the three-dimensional structure of our Galaxy beyond cosmic-ray source distributions. This will result in inhomogeneous transport parameters leading to a less smooth distribution of the cosmic-ray flux in our Galaxy. While this will offer additional constraints, at the same time it should lead to a more precise understanding of cosmic-ray transport. The local cosmic-ray data are not representative for the global Galactic cosmic-ray flux. It is therefore crucial to combine this information with gamma-ray and other data.

\section*{Acknowledgments}
We thank David Maurin for support in using the CRDB cosmic ray database. Most data used in this study were extracted with the help of this database (see \cite{MaurinEtAl2014AnA569A_32}).
Support by the Austrian Federal Ministry of Science, Research and Economy (bmwfw) as part
of the Konjunkturpaket II
of the Focal Point Scientific Computing
 at the University of Innsbruck and the Austrian Science Fund (FWF)  is acknowledged.
M. Werner would like to thank the Max Planck Institute
for Extraterrestrial Physics for its hospitality during his research stay there.
This research has made use of NASA’s Astrophysics Data System. Plots were produced using the matplotlib library (see \cite{Hunter2007CiSE9_90}).


\appendix

\section{Evaluation of numerical accuracy}
\label{SecNumAnalysis}
In the following sections we verify the capability of \textsc{Picard} to produce accurate solutions to 3D Galactic cosmic-ray propagation models. For this we first show that for an axisymmetric setup we recover the same solutions as are found with \textsc{Galprop} using a 2D setup. We illustrate that the axial symmetry is conserved when using our Cartesian grid. We discuss the convergence and accuracy of our numerical results and the impact of multiple network iterations.

\subsection{Comparison of 3D \textsc{Picard} to 2D \textsc{Galprop}}
As in 3D \textsc{Galprop}, \textsc{Picard} uses a Cartesian grid in configuration space, whereas the \textsc{Galprop} 2D models feature a cylindrical grid (\textsc{Galprop} models are prominently obtained using the 2D solver only).
The choice of the numerical grid should locally, at least for high resolution, have no effect on the symmetry of the solution. But because  of the usual $\psi=0$ boundary conditions, the different positions of the domain boundaries in 2D and 3D model can potentially influence the resulting cosmic-ray distribution.

To quantify these possible deviations we solved all model setups given in table \ref{TabParameters} with the 2D \textsc{Galprop} and the 3D \textsc{Picard} solver. For this we used a spatial grid with 129 grid points in the $x$- and $y$-dimension and 33 grid point in the $z$-direction for the 3D simulations and 65 grid points in the $r$-dimension for the 2D models.

 From a direct comparison of the results obtained using the two different solvers we computed relative differences of the models, which show that the different domain boundaries in most models have only negligible influence on the spectra at the location of Earth. Differences of the spectra were found to be on the few-percent level, which would hardly change the quality of a fit to data for any cosmic-ray nucleus. 

\begin{figure}
  \setlength{\unitlength}{0.008cm}
  \centering
  \begin{picture}(1100,858)(-100,-100)
    \put(420,-70){$E$ [GeV]}%
    \put(-70,330){\rotatebox{90}{B/C}}
    \includegraphics[width=1000\unitlength]{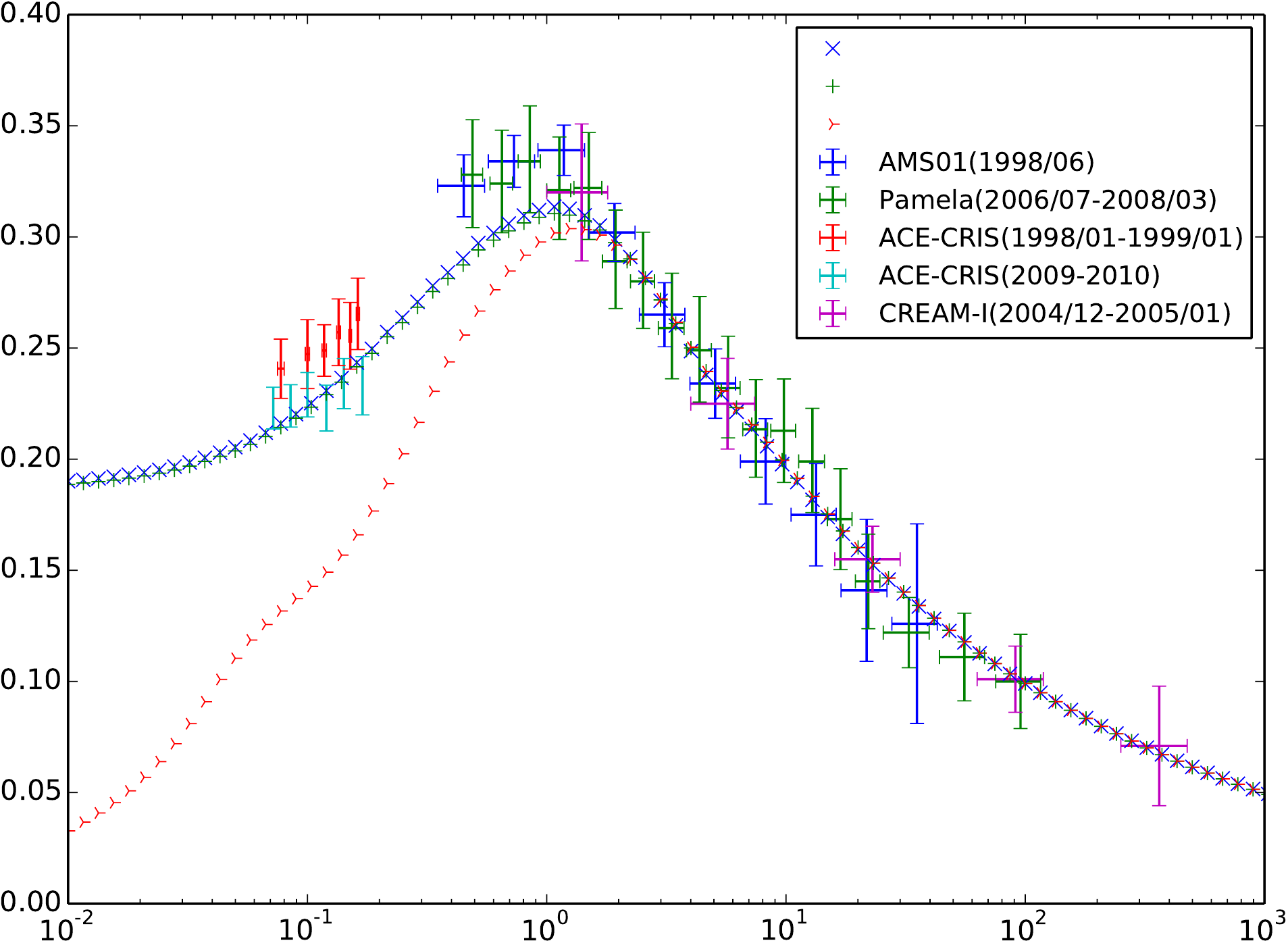}
    \put(-340,688){\tiny 2D \textsc{Galprop}}
    \put(-340,658){\tiny 3D \textsc{Picard}}
    \put(-340,628){\tiny 3D \textsc{Picard} (LIS)}
  \end{picture}
  \caption{\label{FigBCFiducial}B/C ratio for our reference parameter set based on model $^{\text{S}}$Y$^{\text{Z}}$4$^{\text{R}}$30$^{\text{T}}$150$^{\text{C}}$5 from \citet{AckermannEtAl2012ApJ750_3} for both the 2D \textsc{Galprop} model and the 3D \textsc{Picard} model (using a modulation potential of $\Phi=350$\,MV). Note that for the 2D \textsc{Galprop} model different time-integration parameters had to be used than in \citet{AckermannEtAl2012ApJ750_3} (see Appendix \ref{AppendGalpropConvergence}).}
\end{figure}
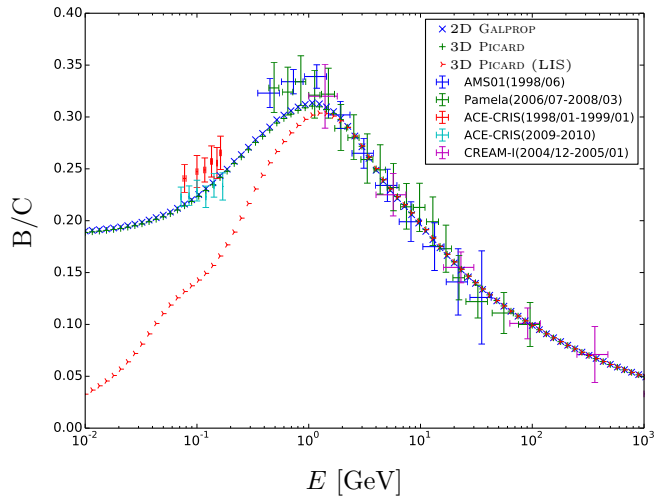

Acknowledging the good accuracy of measurements of B/C, we use this ratio at the nominal position of sun as a proxy for the agreement of the 2D \textsc{Galprop} models and the 3D \textsc{Picard} models. In this comparison we found that the differences were at the 1\% level or below (see Fig. \ref{FigBCFiducial} for an illustration). The only exception was the z8R20 model, where the deviation was  higher, approaching the 2\% level. The small differences found for the z4R20 and the z4R30 models show that there is only negligible influence of the boundary conditions. The remaining difference are probably caused by remaining differences of the numerical solvers.

\subsubsection{Convergence of \textsc{Galprop} models}
\label{AppendGalpropConvergence}
In the foregoing comparison, we found that the reference parameter sets taken from \citet{AckermannEtAl2012ApJ750_3} did not show a fully converged steady state solution. We only found a satisfying agreement between our 3D models and the 2D \textsc{Galprop} models, after adapting the time-integration parameters. This shows that for some models in the literature the convergence of the numerical solution was apparently not verified.

In contrast to the solver used in \textsc{Picard}, the \textsc{Galprop} solver uses a time-integration procedure to find a steady state solution of the transport equations. Therefore, it is necessary to verify that the solver has indeed found such a steady state solution and adapt the time-integration procedure if this is not the case (see \cite{WernerEtAl2013arXiv1308_2829W}).

\begin{figure*}
  \setlength{\unitlength}{0.00044\textwidth}
  \begin{picture}(1100,874)(-100,-100)
    \put(360,-70){\small$E$ [GeV]}
    \put(-70,120){\small\rotatebox{90}{relative deviation [\%]}}
    \includegraphics[width=1000\unitlength]{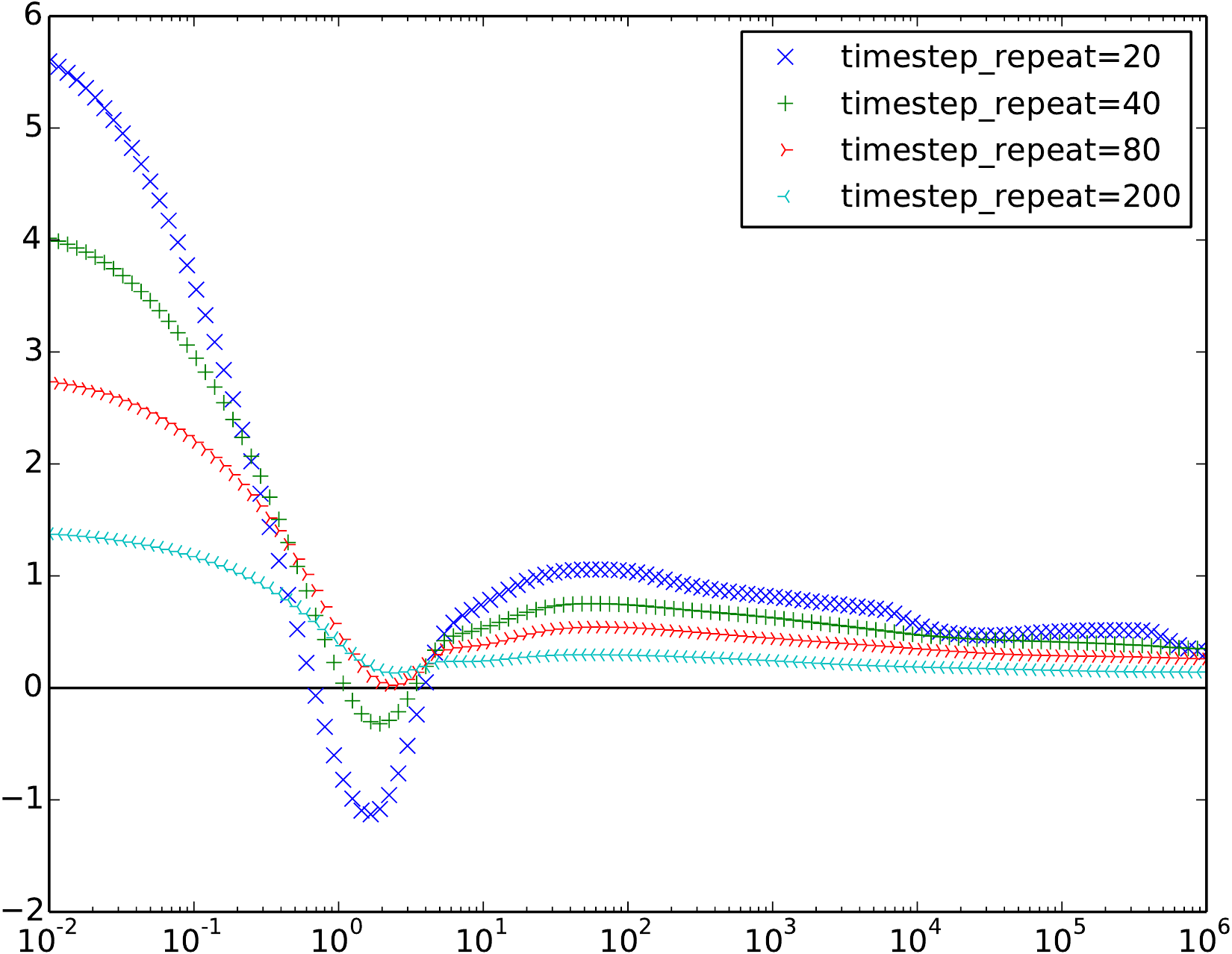}
  \end{picture}
  \hfill
  \begin{picture}(1000,874)(0,-100)
    \put(360,-70){\small$E$ [GeV]}
    \includegraphics[width=1000\unitlength]{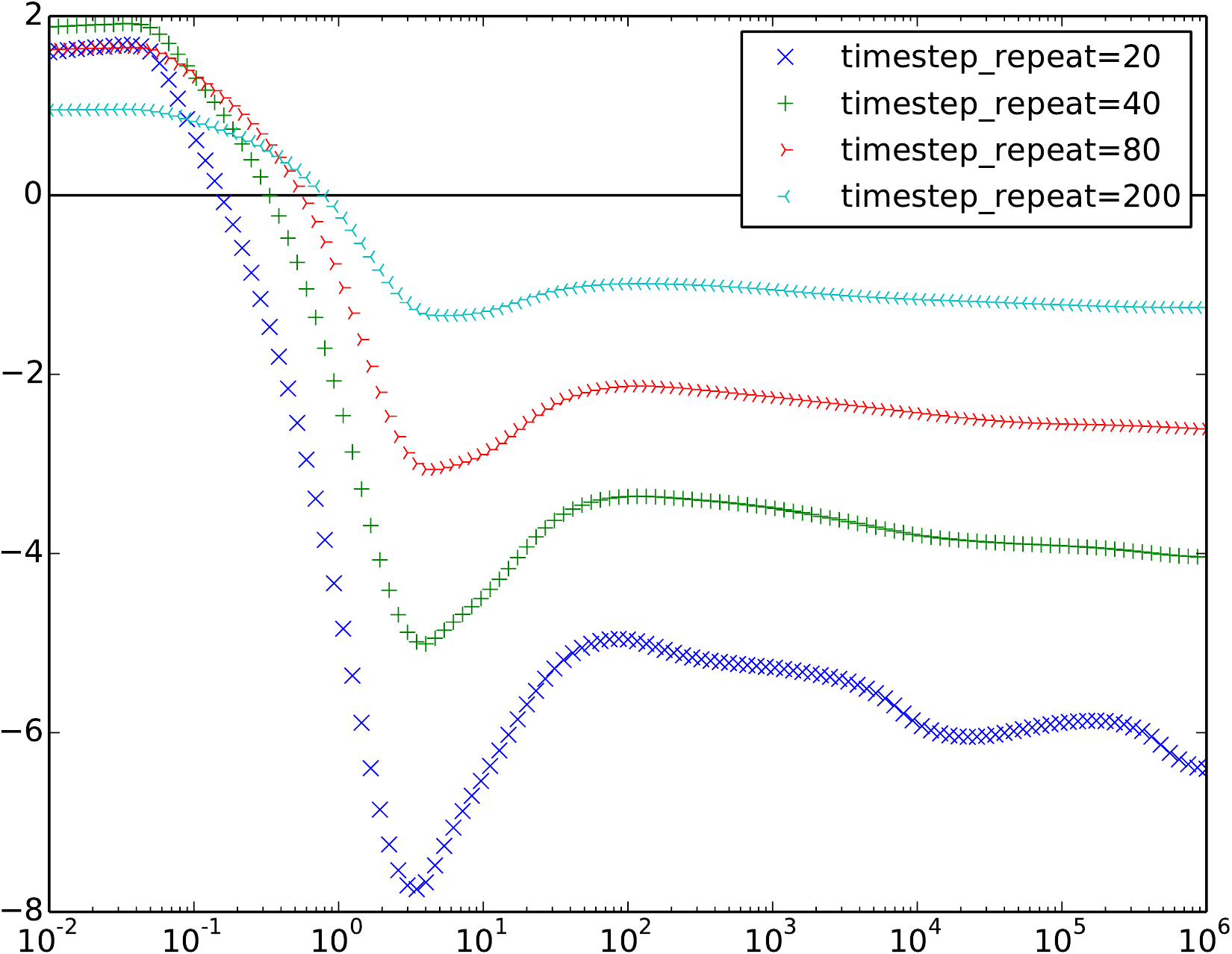}
  \end{picture}
  \caption{\label{FigGalpropConverge} Illustration of the relative deviation of \textsc{Galprop} models using different time-integration parameters. On the left the $^{12}$C spectrum and on the right the $^{10}$B spectrum near Earth are shown. Deviations are given in percent relative to a solution computed with \texttt{timestep\symbol{95}repeat}=1000. Simulation parameters are adapted from model $^{\text{S}}$Y$^{\text{Z}}$4$^{\text{R}}$30$^{\text{T}}$150$^{\text{C}}$5 from \citet{AckermannEtAl2012ApJ750_3}.}
\end{figure*}

We found that the time-integration parameters in our 2D \textsc{Galprop} reference models taken from \citet{AckermannEtAl2012ApJ750_3} did not lead to a converged steady state solution. Correspondingly, we conducted a brief study to determine the influence of the time-integration parameters on the final solution. We found that the most sensitive parameter in a \textsc{Galdef} file seems to be \verb|timestep_repeat|, i.e., the number of times a certain time step is to be repeated by the solver. By increasing this number we found that the computed fluxes at the position of the sun could change by several percent as illustrated in Fig. \ref{FigGalpropConverge}.

With the standard setup using a parameter \verb|timestep_repeat|=20 we find deviations from the presumed steady-state solution up to and possibly exceeding 8\% (see Fig. \ref{FigGalpropConverge}). Since the energy-dependence of this deviation differs markedly for primaries and secondaries, it will also be relevant for the resulting flux ratios. A deviation of near 10\% for the flux ratios, however, can significantly decrease the fit quality to current data.

This discussion is meant to caution that whenever a set of propagation parameters is cited,  one has to test the validity for the problem at hand. In the present study we only found
near convergence to the steady state when \verb|timestep_repeat| was set to a value of several hundred. In that case, however, even the 2D solver in \textsc{Galprop} is at pair with the 3D solver in \textsc{Picard}.

\subsection{Testing Axial Symmetry}
\label{SecAppendAxial}
As a further test we investigated the axial symmetry of the models computed with the 3D \textsc{Picard} solver directly. For this, we compared spectra at $\phi=0$ and $\phi=45$ in the Galactic plane, where the former direction coincides with the Cartesian $x$-axis and the direction to the sun and the latter is in the direction of a corner of the computational domain. Thus, the largest differences are to be expected in these directions.
\begin{figure*}
  \setlength{\unitlength}{0.00044\textwidth}
  \begin{picture}(1100,874)(-100,-100)
    \put(390,-70){\small$E$ [GeV]}
    \put(-70,120){\small\rotatebox{90}{relative deviation [\%]}}
    \includegraphics[width=1000\unitlength]{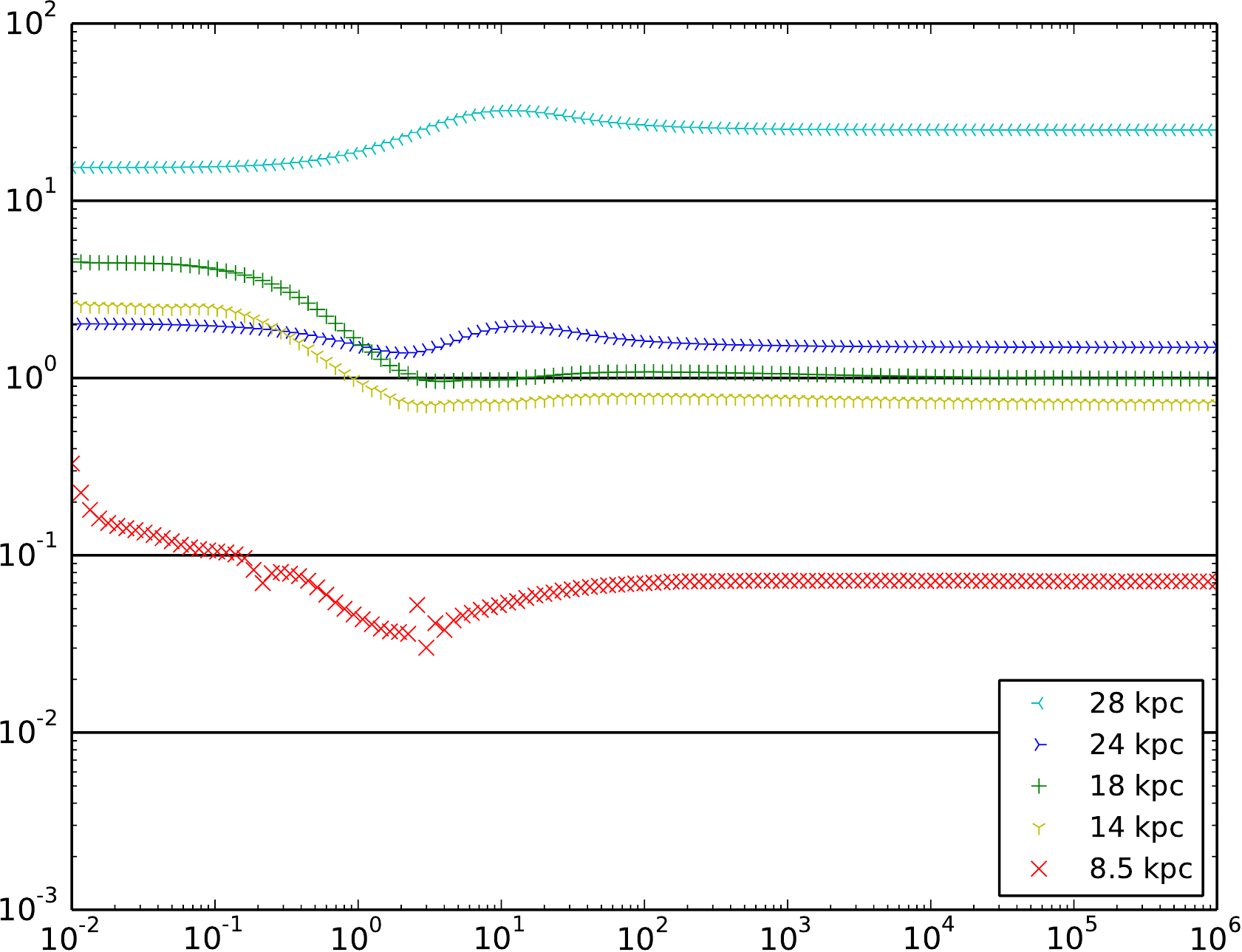}
  \end{picture}
  \hfill
  \begin{picture}(1000,874)(0,-100)
    \put(390,-70){\small$E$ [GeV]}
    \includegraphics[width=1000\unitlength]{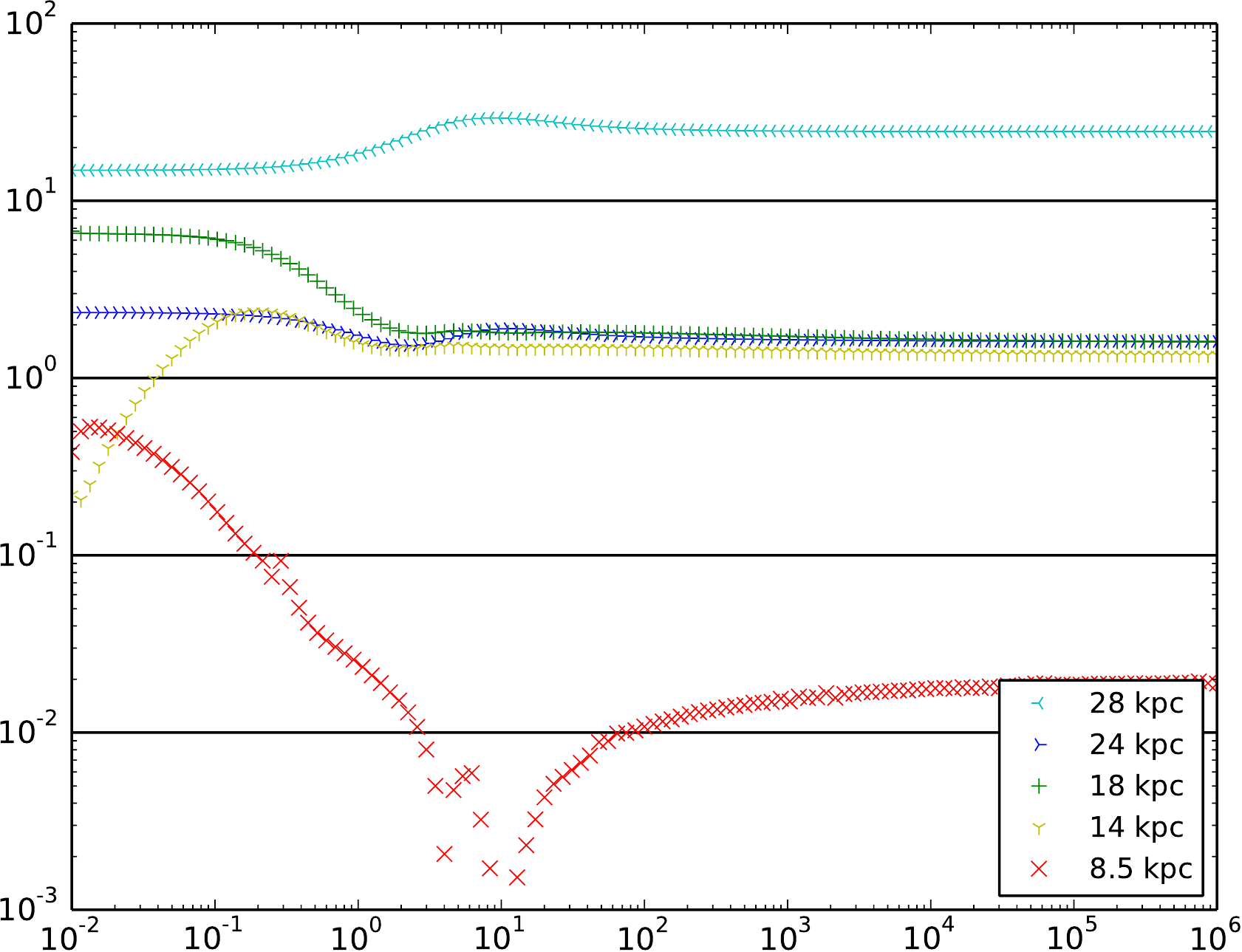}
  \end{picture}
  \begin{picture}(1100,874)(-100,-100)
    \put(390,-70){\small$E$ [GeV]}
    \put(-70,120){\small\rotatebox{90}{relative deviation [\%]}}
    \includegraphics[width=1000\unitlength]{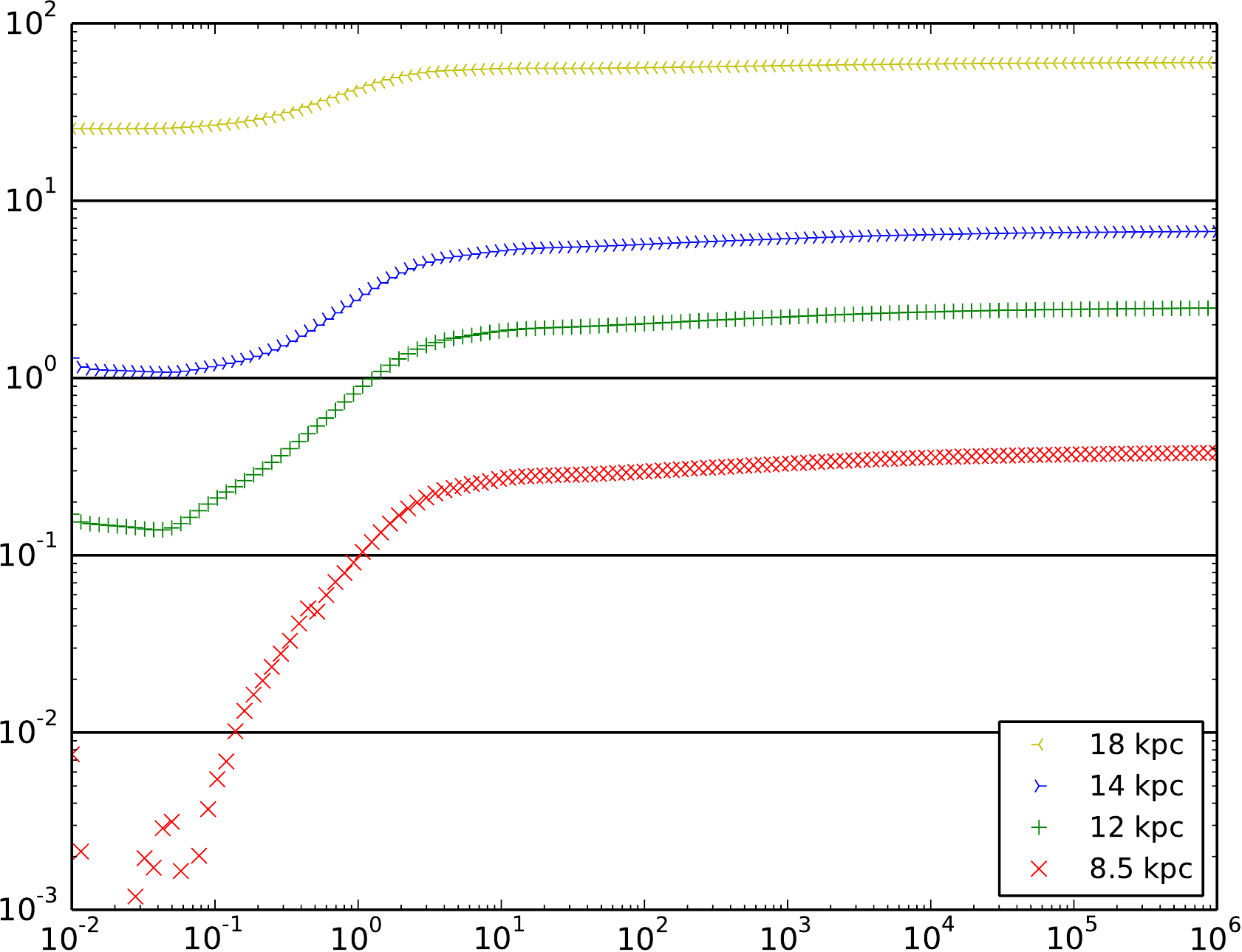}
  \end{picture}
  \hfill
  \begin{picture}(1000,874)(0,-100)
    \put(390,-70){\small$E$ [GeV]}
    \includegraphics[width=1000\unitlength]{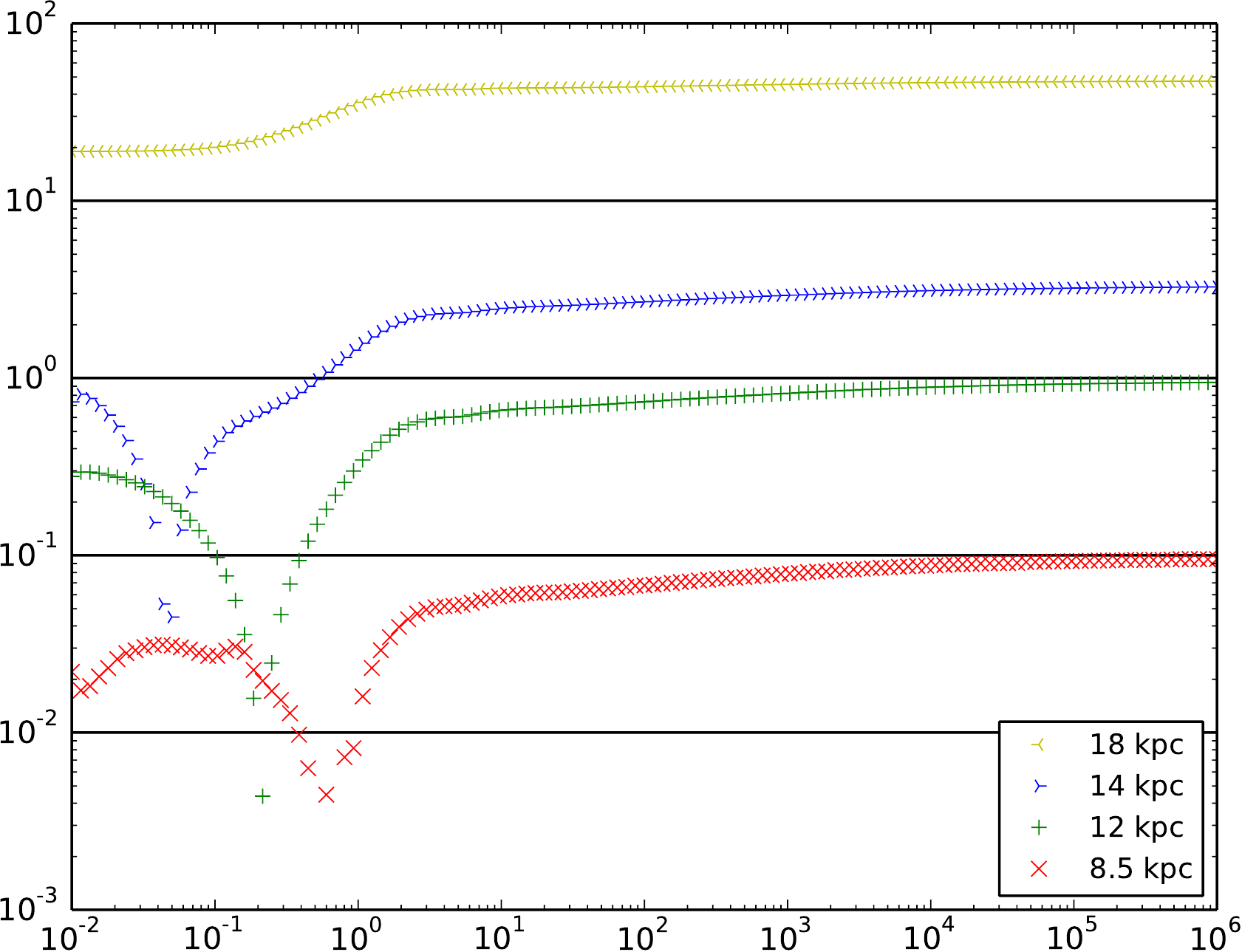}
  \end{picture}
  \caption{\label{FigCompAxiSymm} Relative deviation of the spectra in the Galactic plane at $\phi=45$ from those at $\phi=0$ for the same galactocentric distance. Results are shown for different radii as indicated in the figures. Here, we show relative deviations for the 3D \textsc{Picard} solution of models z8R20 (bottom) and z4R30 (top); on the left results are shown for $^{10}$B and on the right for $^{12}$C.}
\end{figure*}

We found that the differences in the cosmic-ray spectra at a Galactocentric radius of $r=8.5$\,kpc are below the percent level (see Fig. \ref{FigCompAxiSymm} for some examples). Only for larger distances from the Galactic centre do we observe an increasing deviation from axial symmetry, which only becomes significant in the vicinity of the radial boundary of the domain.

For models with a Galactic radius of 20\,kpc the deviations turn out to be noticeably higher for secondaries than for primaries (see bottom panels in Fig. \ref{FigCompAxiSymm}). This difference increases with increasing halo height. In the large-scale simulations with $x,y = -30\dots30$\,kpc, in contrast to that, the behaviour for primaries and secondaries is much more similar (see top panels in Fig. \ref{FigCompAxiSymm}). While in the 20\,kpc simulations deviations systematically increase with increasing halo height, this only happens near the outer radial boundary for the 30\,kpc simulations.

The different behaviour of secondaries and primaries for the case of a smaller radial domain is caused by the fact that primaries trace the cosmic-ray source distribution much closer than secondaries do, as was also found for the spiral-arm models. With the cosmic-ray source distribution going to zero at $r=15\,kpc$ the influence of the boundary is smaller for the primaries than for the secondaries. For the large scale simulations in contrast, the domain boundary is so far from the cosmic-ray sources that there is not much difference between primaries and secondaries in any of the simulations. This can also be concluded from the observation that the error becomes nearly distance-independent in the range from 14-24\,kpc in those models with a large radial extent.

From the foregoing discussion we conclude that the influence of the outer radial boundary of the numerical domain on the distribution of cosmic rays in the inner part of the domain depends on both the radial size of the numerical domain and also on the halo height. Especially the simulation with a large halo height and a small radial extent, i.e. simulation z8R20, indicates some effects of the boundary conditions even in the inner parts of the Galaxy. Together with the higher deviation of B/C found in the previous section, this shows that model z8R20 behaves problematicly in the 3D Cartesian setup. All other models can be used to extend previous axisymmetric 2D models to spatially 3D propagation problems.

\subsection{Accuracy of the Numerical Solution}

In the following we quantify the accuracy of the numerical solution computed with \textsc{Picard}. For this we investigate the error resulting from the discretisation of the transport equation in configuration and momentum space. Additionally, we determine the impact of the number of network iterations on the resulting cosmic-ray flux for the different nuclei.

\subsubsection{Resolution study}
\label{AppendResStudy}

As a reference model for the investigation of the momentum discretisation error we solved model z4R20 with an axisymmetric source distribution using $N_{x,y}$=129 grid points in $x$ and $y$ and $N_z$=33 grid points in $z$ together with 255 logarithmically equidistant grid points in momentum space. Results of this model were then compared to a low-resolution model with 63 grid points and a mid-resolution model with 127 grid points in momentum space. We found that the average difference between the low- and the mid-resolution setup already is as low as 2\% for the $^{12}$C spectrum at Earth.

For the investigation of the spatial discretisation error we only varied the spatial resolution using 127 logarithmically equidistant grid points in momentum space. As the reference solution we used a simulation with $N_{x,y}=$257 grid points in $x$ and $y$ and $N_z=$65 grid points in $z$.  This high-resolution model was then compared to a low-resolution simulation ($N_{x,y}$ = 65; $N_z$ = 17) and a mid-resolution simulation ($N_{x,y}$ = 129; $N_z$ = 33).

We found an average deviation of some 4.4\% for the low-resolution simulation and of 2\% for the mid-resolution simulation relative to the reference model. 
This spatial discretisation error turns out to be dominated by the discretisation in the $z$-dimension, also reflecting the usual choice of a resolution of 0.1\,kpc in $z$ and 1\,kpc in the radial direction common in most \textsc{Galprop} simulations. This is caused by the sharp decrease of the cosmic-ray source function perpendicular to the Galactic plane in this axisymmetric setup.
Since the spiral-arm source distributions also have strong gradients in the radial direction, however, we use a numerical grid with approximately the same resolution in all spatial dimensions. Motivated by the present study, we use a setup with $N_{x,y}$=257, $N_z$=65 together with 127 grid points in momentum space throughout this work, if not indicated otherwise.

\subsubsection{Network iterations}
\label{AppNetIter}
To determine the effect of multiple network iterations we repeated model z4R20 with different numbers of network iterations, using $N_{x,y}$=129, $N_z$=33 and 127 grid points in momentum space. By using a simulation with five network iterations as our reference, we found by a comparison of the spectra for different nuclei that only the spectra of
such nuclei that are in a mass range of the nuclear network where secondaries are produced that occur earlier in the network, are  severely affected by a different number of network iterations.

\begin{figure*}
  \setlength{\unitlength}{0.00044\textwidth}
  \begin{picture}(1100,866)(-100,-100)
    \put(330,-70){\small$E_{kin}$/nuc [GeV]}
    \put(-70,-20){\small\rotatebox{90}{flux, m$-^2$\,s$^{-1}$\,sr$^{-1}$\,(GeV/nuc)$^{-1}$}}
    \includegraphics[width=1000\unitlength]{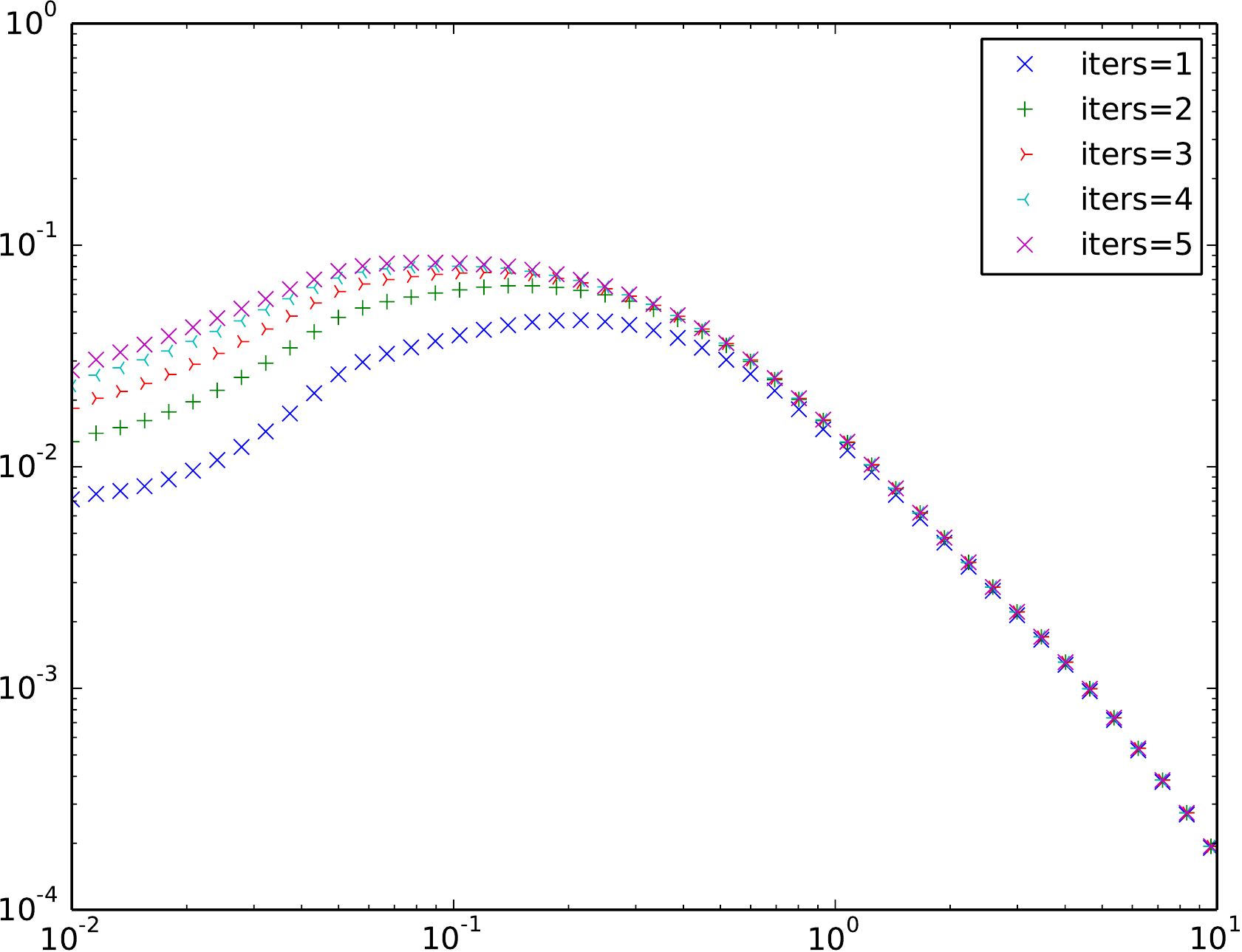}
    \put(-900,650){$^{53}$Mn}
  \end{picture}
  \hfill
  \begin{picture}(1100,866)(-100,-100)
    \put(330,-70){\small$E_{kin}$/nuc [GeV]}
    \put(-70,-20){\small\rotatebox{90}{$E^2$flux,  (GeV/nuc)\,m$-^2$\,s$^{-1}$\,sr$^{-1}$}}
    \includegraphics[width=1000\unitlength]{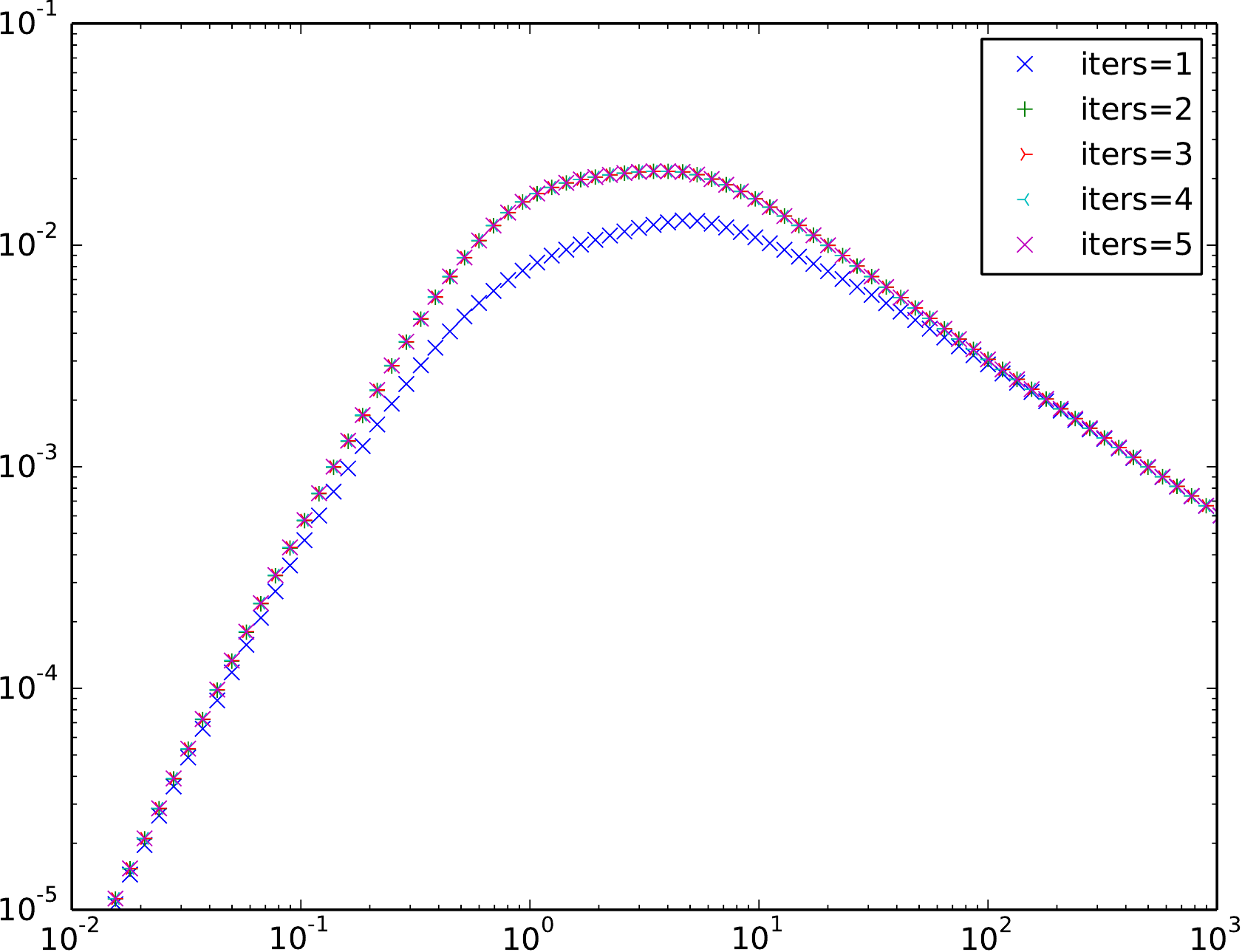}
    \put(-900,650){$^{36}$Ar}
  \end{picture}
  \caption{\label{FigSpecNetworkIters}Resulting cosmic-ray spectra for different numbers of network iterations.
  }
\end{figure*}
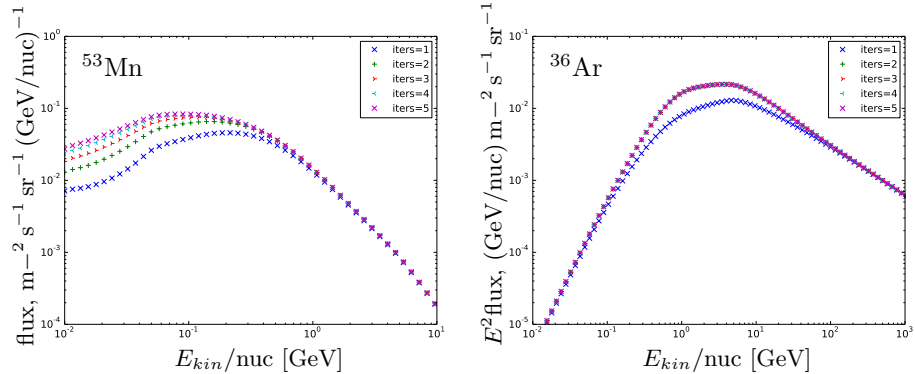



In Fig. \ref{FigSpecNetworkIters} we show two of the most extreme cases, where we compare the resulting spectra using different numbers of network iterations. Apparently, the deviations are most severe at very low energies, where heliospheric modulation makes an interpretation of the results problematic anyhow.  Deviations up to some 50\% in the case of $^{36}$Ar and even exceeding 70\% for $^{53}$Mn for a single network iteration are significantly too large when being used for a quantitative analysis of these nuclei.

For the majority of other cosmic-ray nuclei the effect of multiple network iterations
is negligible, resulting in changes well below 1\%.
Even such important secondary-to-primary ratios as B/C and $^{10}$Be/$^{9}$B show only deviations at -- or in most cases even below -- the 0.1 percent level, when at least two network iterations are used.
This shows that in many applications two network iterations are sufficient for an accurate solution.
For the important case of diffuse gamma-ray emission resulting from the interaction of cosmic rays with the interstellar medium (see, e.g. \cite{AckermannEtAl2012ApJ750_3}) even a single network iteration is sufficient since hydrogen and helium show a negligible change when using multiple network iterations.

\bibliography{$HOME/LaTeX/Bibliographies/GalacticCR,$HOME/LaTeX/Bibliographies/pubrk.bib}
\end{document}